\documentclass[12pt]{article}
\usepackage{amsmath}
\usepackage{enumerate}
\usepackage{amsfonts}

\begin{document}
\def\be{\begin{equation}}
\def\bea{\begin{eqnarray}}
\def\ee{\end{equation}}
\def\eea{\end{eqnarray}}
\def\d{\partial}
\def\eps{\varepsilon}
\def\la{\lambda}
\def\b{\bigskip}
\def\nn{\nonumber \\}
\def\t{\tilde}
\def\k{\kappa}
\def\eps{\varepsilon}
\def\ch{\mbox{ch}}
\def\sh{\mbox{sh}}

\makeatletter
\def\blfootnote{\xdef\@thefnmark{}\@footnotetext}  
\makeatother

\begin{center}
{\LARGE Generalized $\la$--deformations of AdS$_p\times$S$^p$}
\\
\vspace{18mm}
{\bf   Yuri Chervonyi  and   Oleg Lunin}
\vspace{14mm}

Department of Physics,\\ University at Albany (SUNY),\\ Albany, NY 12222, USA\\ 

\vskip 10 mm

\blfootnote{ichervonyi@albany.edu,~olunin@albany.edu}

\end{center}

\begin{abstract}

We study analytical properties of the generalized $\la$--deformation, which modifies string theories while preserving integrability, and construct the explicit backgrounds corresponding to AdS$_p\times$S$^p$, including the Ramond--Ramond fluxes. For an arbitrary coset, we find the general form of the R--matrix underlying the deformation, and prove that the dilaton is not modified by the deformation, while the frames are multiplied by a constant matrix. Our explicit solutions describe families of integrable string theories depending on several continuous parameters.

\b

\end{abstract}

\newpage

\tableofcontents

\section{Introduction}

The last few years witnessed an impressive progress in finding new families of integrable string theories. Initially integrability was discovered in isolated models, such as strings on AdS$_p\times$S$^q$ \cite{IntAdS5,AdS3Int,IntAdS2}, and in their extensions called beta deformations \cite{beta}. Recent developments, stimulated by the mathematical literature \cite{Cherednyk}, led to construction of very large classes of integrable string theories. One of the approaches originated from studies of the Yang--Baxter sigma models \cite{BosonicYangBaxter,Qdeform,Yoshida}, and it culminated in construction of new integrable string theories, which became known as $\eta$--deformations \cite{Delduc1,OtherEta,MoreEta}.
A different approach originated from the desire to relate two classes of solvable systems, the Wess--Zumino--Witten \cite{WZW} and the Principal Chiral \cite{PCM} sigma models, and it culminated in the discovery of a one--parameter family of integrable conformal field theories, which has WZW and PCM as its endpoints \cite{SftsGr,ST}\footnote{See \cite{PreST,PreST1} for earlier work in this direction.}. Such line of conformal field theories becomes especially interesting when the PCM point represents a string theory on AdS$_p\times$S$^q$ space, and the corresponding families, which became known as $\la$--deformations, have been subjects of intensive investigations \cite{SfetsosAdS5,HT,HMS,BTW,Holl2}. Recently the powers of the two approaches were combined to construct the generalized $\la$--deformations \cite{GenLambda}\footnote{See 
\cite{HT} for the earlier exploration of the connection between the $\eta$ and $\la$ deformations.}, the largest class on integrable string theories known to date, which encompasses all earlier examples.  In this article we study the generalized $\la$--deformations of cosets with a special emphasis on describing integrable extensions of strings on AdS$_2\times$S$^2$, AdS$_3\times$S$^3$, and AdS$_5\times$S$^5$.
 
\bigskip

While the procedure for constructing the generalized $\la$--deformation has been outlined in \cite{GenLambda}, its practical implementation presents some technical challenges. Moreover, just as in the case of the standard $\la$-- and $\eta$--deformations, the CFT construction gives only the NS--NS fields, and evaluation of the Ramond--Ramond fluxes relies on supergravity computations. On the CFT side one encounters two types of challenges: construction of the classical R--matrix, which is the central element of the generalized $\la$--deformation, and evaluation of the modified metric. R--matrices are solutions of the modified classical Yang--Baxter equation (mCYB), and while many examples have been studied in the literature \cite{CYB, BosonicYangBaxter}, the full classification of R--matrices is still missing. In section \ref{Rmatrix} we find a rather general class of solutions of the mCYB equation for arbitrary cosets $G/F$, and for specific examples arising in the description of strings on AdS$_p\times$S$^p$ we construct {\it all} solutions. Keeping in mind that the prescription of \cite{GenLambda} might have a counterpart involving supercosets (as it happened in the case of the ordinary $\la$--deformation \cite{HMS,BTW,ChLLambda}), we also find a large class of R--matrices solving the graded mCYB equation, which governs the deformations of supercosets. Deforming various supercosets using such matrices would be an interesting topic for future work.

Finding the R--matrices is not the only technical challenge associated with the generalized $\la$--deformation. 
While the procedure for finding the metric is algorithmic, and in principle it can be applied to any coset\footnote{In practice, the difficulty of such `brute force' calculation grows exponentially with the size of the coset and the number of deformation parameters. This presents an additional motivation for understanding the hidden symmetries of the problem and for simplifying the calculations.}, the calculations can be tedious, and one finds a lot of `accidental cancellations' in the final results. Such surprises have been encountered in the past \cite{ST, SfetsosAdS5}, and in some instances they have been explained on a case-by-case basis \cite{SfetsosAdS5}. In section \ref{SUGRAemb} we demonstrate that the `accidental cancellations' are guaranteed by the symmetries of the underlying problem, thus they must be present for all deformations, and they can be used to drastically simplify the calculations. 
Even apart from this practical usefulness, our study of hidden symmetries contributes to the general 
analytical understanding of integrable deformations. 

Application of the algebraic procedure outlined in \cite{GenLambda} yields the metric and the dilaton for the deformed backgrounds, but recovery of the Ramond--Ramond fluxes from the sigma model is a very complicated task \cite{BTW}. In practice, it is much easier to find such fluxes by solving the supergravity equations of motion, and in the past this technique has been successfully implemented for several families of integrable string theories \cite{OtherEta,ST, SfetsosAdS5, ChLLambda}. Following the same path in section \ref{SUGRAemb}, we recover the fluxes supporting the generalized $\la$--deformation of  AdS$_2\times$S$^2$ and AdS$_3\times$S$^3$. Interestingly, the construction of \cite{GenLambda} does not allow one to deform AdS$_5\times$S$^5$ unless a trivial R--matrix is chosen.

\bigskip

This paper has the following organization. In section \ref{SectReview} we review the procedure for finding the generalized $\la$--deformation introduced in \cite{GenLambda}. This construction is based on solutions of the classical modified Yang--Baxter equation, and in section \ref{Rmatrix} we find large classes of such solutions for general cosets $G/F$, as well as the most general solutions that can be used to deform string theory on  AdS$_p\times$S$^p$ ($p=2,3,5$). We also construct very large classes of graded R--matrices, which can be used for extending the procedure of \cite{GenLambda} to supercosets, along the lines of the analysis presented in \cite{HMS}. In section \ref{SubsGenMetr} we uncover some analytical properties of the deformed metric and the dilaton, which are applicable to all cosets. The remainder of section \ref{SUGRAemb} is devoted to constructing the supergravity backgrounds supporting the generalized $\la$--deformations of  AdS$_p\times$S$^p$. Appendix \ref{AppDmtr} is devoted to exploration of analytical properties of a matrix that plays a pivotal role in constructing the generalized $\la$--deformations.

\section{Review of the generalized $\lambda$-deformation}
\label{SectReview}
\renewcommand{\theequation}{2.\arabic{equation}}
\setcounter{equation}{0}

Lambda deformations of the Principal Chiral Models (PCM) were introduced in \cite{SftsGr} and further studied in \cite{HMS, ST, SfetsosAdS5, HT, BTW, ChLLambda}. Application of such deformation to any PCM leads to a one--parameter family of integrable conformal field theories. This deformation was generalized to a larger family in \cite{GenLambda}, and we begin with reviewing this construction following section 5 of \cite{GenLambda}.

\bigskip

The $\lambda$ deformation  interpolated between Conformal Field Theories described by a Principal Chiral Model (PCM) and a Wess--Zumino--Witten model (WZW), and we begin with looking at the WZW side:
\bea\label{gWZW}
S_{WZW,k}(g)=\frac{k}{4\pi}\int_\Sigma d^2\sigma R_+^a R_-^a-\frac{k}{24\pi}\int_B f_{abc} R^a\wedge R^b\wedge R^c,\quad \d B=\Sigma.
\eea
Here $g\in G$ is an element of some group $G$ with generators $T_a$, $k$ is the level of the WZW model, 
$R_\pm$ are the right-invariant Maurer-Cartan forms,
\bea
R^a_\pm=-i \mbox{Tr}(T^a \d_\pm g  g^{-1})\,,
\eea
and $f_{abc}$ are the structure constants:
\bea
[T_a,T_b]=i f_{ab}{}^c T_c\,.
\eea
To construct the $\lambda$ deformation one adds the action (\ref{gWZW}) to a generalized PCM on a group manifold\footnote{In comparison with \cite{GenLambda} we have rescaled the constant coefficients ${E}_{ab}$ by $k$ so the level of the WZW appears as an overall factor in the sum of (\ref{gWZW}) and (\ref{genPCM}). Such rescaling simplifies the formulas associated with $\la$--deformation.},
\bea\label{genPCM}
S_{gPCM}(\hat{g})=\frac{k}{2\pi}\int d^2\sigma {E}_{ab} R^a_+(\hat{g}) R^b_-(\hat{g}),\qquad
\hat{g}\in G\,,
\eea
and gauges away half of the degrees of freedom in the resulting 
sum\footnote{See \cite{GenLambda} for more details.}.
Parameters ${E}_{ab}$ in (\ref{genPCM}) represent an arbitrary constant matrix, and later its form will be restricted by the requirements of conformal invariance and integrability. The gauging procedure in the sum of (\ref{gWZW}) and (\ref{genPCM}) leads to the action \cite{PreST1,GenLambda}
\bea\label{GenLambdaAction}
S_{k,\lambda}(g)=S_{WZW,k}(g)+\frac{k}{2\pi}\int d^2\sigma L_+^a(\hat{\lambda}^{-1}-D)^{-1}R_-^b,
\eea
where\footnote{Following \cite{GenLambda}, we denote the {\it matrix} appearing in (\ref{GenLambdaAction}), (\ref{lambdaMtrDef}) by $\hat{\lambda}$ to distinguish it from the {\it scalar} deformation parameter $\la$.}  
\bea\label{lambdaMtrDef}
\hat{\lambda}^{-1}={E}+ I,
\quad D_{ab}=\mbox{Tr}(T_a g T_b g^{-1}),\quad
L_\pm^a
=i\mbox{Tr}(T_ag^{-1}\d_\pm g),\quad R_\mu^a=D_{ab}L_\mu^b.
\eea
Application of this prescription to the standard PCM,
\bea\label{StandLambda}
{E}_{ab}=\frac{\kappa^2}{k}\delta_{ab},\quad \hat{\lambda}^{-1}=\frac{k+\kappa^2}{k}I,
\eea
leads to a one-parameter $\lambda$--deformation, and integrability of the corresponding conformal field theory (\ref{GenLambdaAction}) was demonstrated in \cite{SftsGr}. It is clear that the sigma model \eqref{GenLambdaAction} would not be integrable for a generic matrix ${E}$, but the authors of \cite{GenLambda} found a large class of integrable models extending (\ref{StandLambda}). We begin with reviewing this construction for groups, and then discuss the cosets, which will be the main objects of our study.

\bigskip
\noindent
{\bf Generalized $\lambda$-deformation for groups.}

To arrive at an integrable deformation (\ref{GenLambdaAction}), one should start with an integrable generalized PCM (\ref{genPCM}), and this already imposes severe restrictions on the constant matrix 
${ E}_{ab}$. Extending the standard choice (\ref{StandLambda}), one can start with the action of the $\eta$--deformed PCM \cite{BosonicYangBaxter}:
\bea\label{YangBaxterAction}
S_{gPCM}=\frac{1}{2\pi {\tilde t}}\int d^2\sigma R_+^T(I-\tilde{\eta} \mathcal{R})^{-1}R_-,\quad \eta>0.
\eea
As demonstrated in \cite{BosonicYangBaxter}, this model is integrable, as long as the constant matrix $\mathcal{R}$ satisfies the modified classical Yang-Baxter (mCYB) equation\footnote{The constant matrix $\mathcal{R}$ satisfying the Yang-Baxter equation is called the Yang-Baxter operator or the R--matrix. In this paper we use both names.}
\bea\label{modYBeq}
[\mathcal{R} A,\mathcal{R} B]-\mathcal{R}([\mathcal{R} A,B]+[A,\mathcal{R} B])=-c^2 [A,B],\quad A,B\in \mathfrak{g},\quad c\in\mathbb{C}.
\eea
Then the interpolating model (\ref{GenLambdaAction}) with 
\bea
{E}_{YB}=\frac{1}{\tilde{t}}\left( I-\tilde{\eta}\mathcal{R} \right)^{-1}
\eea
is integrable as well, and it is called the generalized $\lambda$-deformation of (\ref{YangBaxterAction}) \cite{GenLambda}.

\bigskip
\noindent
{\bf Generalized $\lambda$-deformation for cosets.}

The authors of \cite{GenLambda} also extended the construction of the generalized $\lambda$-deformation to cosets $G/F$ by defining 
\bea\label{EmtrCosetA}
{E}={E}_H\oplus{E}_{G/F},\quad {E}_F=0,\quad {E}_{G/F}=\frac{1}{\tilde{t}}(I-\tilde{\eta}\mathcal{R})^{-1},\quad \mathfrak{g}=\mathfrak{f}+\mathfrak{l},
\eea
This ansatz for ${E}$ leads to inconsistent equations of motion for (\ref{GenLambdaAction}) unless all elements of the coset satisfy the constraint \cite{GenLambda}\footnote{This constraint is multiplied by $\tilde\eta$, but since we are interested in the deformed theory, $\tilde\eta\ne 0$}:
\bea\label{YBcosetConstr}
([\mathcal{R}X,Y]+[X,\mathcal{R}Y])|_\mathfrak{f}=0,\quad X,Y\in\mathfrak{l}.
\eea
Assuming that this constraint is satisfied, the equations of motion for the action \eqref{GenLambdaAction} with the matrix ${E}$ from \eqref{EmtrCosetA} can be written as the integrability condition of a Lax pair (see \cite{GenLambda} for details).

\bigskip

To summarize, the generalized $\lambda$ deformation can be defined on cosets, but integrability puts a severe restriction (\ref{YBcosetConstr}) on the Yang-Baxter operator $\mathcal{R}$. In the next section we will consider several cosets arising in the type II string theory and discuss the corresponding Yang-Baxter operators $\mathcal{R}$ solving the modified classical Yang-Baxter (mCYB) equation \eqref{modYBeq} and the coset constraint \eqref{YBcosetConstr}. Then in section \ref{SUGRAemb} we will use these solutions to embed the generalized $\lambda$ deformations of the corresponding cosets into supergravity.


\section{R-matrices for Lie algebras and cosets}\label{Rmatrix}
\renewcommand{\theequation}{3.\arabic{equation}}
\setcounter{equation}{0}

In string theory integrability was discovered by studying strings on AdS$_p\times$S$^q$ \cite{IntAdS5,AdS3Int,IntAdS2} and the corresponding CFTs are the Principal Chiral models on various cosets. In this article we are interested in the generalized $\lambda$ deformations of such backgrounds, so as outlined in the last section, we should find the Yang--Baxter operators $\mathcal{R}$ satisfying the mCYB equation (\ref{modYBeq}) and the constraint (\ref{YBcosetConstr}) on the relevant coset. In subsection \ref{SubsRgen} we will discuss some general features of such operators, and in the remaining part of this section we will apply this construction to the specific cosets arising in string theory.

\subsection{General construction}
\label{SubsRgen}

The generalized $\la$ deformation reviewed in section \ref{SectReview} is based on the Yang-Baxter operator satisfying the mCYB equation (\ref{modYBeq})\footnote{We set $c=i$ in \eqref{modYBeq}.},
\bea\label{mYBeq}
\label{YBmatrix}
[\mathcal{R}X,\mathcal{R}Y]-\mathcal{R}([\mathcal{R}X,Y]+[X,\mathcal{R}Y])=[X,Y],\quad X,Y\in \mathfrak{g},
\eea
and the constraint (\ref{YBcosetConstr})
\bea\label{YBcosetConstr1}
\label{CosetConstrMtr}
([\mathcal{R}\tilde{X},\tilde{Y}]+[\tilde{X},\mathcal{R}\tilde{Y}])|_\mathfrak{f}=0,\quad \mathfrak{g}=\mathfrak{f}+\mathfrak{l},\quad \tilde{X},\tilde{Y}\in\mathfrak{l}\,.
\eea
We further impose the skew-symmetry condition
\bea\label{skewSym}
(\mathcal{R}X,Y)_{\mathfrak{g}}+(X,\mathcal{R}Y)_{\mathfrak{g}}=0,
\eea
where $(.,.)_\mathfrak{g}$ is the Killing-Cartan form on the Lie algebra. While acting on generators $T_a$, the operator $\mathcal{R}$ can be viewed as a tensor with one lower and one upper index (${\mathcal{R}_b}^a$) and the skew-symmetry condition \eqref{skewSym} means that
\bea\label{skewSymMatr}
\mathcal{R}_{ab}=-\mathcal{R}_{ba}\,.
\eea

Finding the most general solution of (\ref{mYBeq}) for an arbitrary group is an open problem, but one solution is well-known \cite{BosonicYangBaxter}, and now we will introduce its generalization. We will also find the most general solution of (\ref{mYBeq})--(\ref{skewSym}) for specific cosets arising in string theory. 

\bigskip

Equations (\ref{mYBeq}), (\ref{skewSymMatr}) in the adjoint representation imply that $\mathcal{R}_{ab}$ is a real antisymmetric matrix, so it can be diagonalized using a {\it unitary} rotation, and all its eigenvalues are imaginary. In particular, some of these eigenvalues might vanish, then equation (\ref{mYBeq}) implies that the corresponding eigenvectors (which are generators of $\mathfrak{g}$) must commute. Thus we conclude that the kernel of operator ${\mathcal{R}_b}^a$ is a subset of the Cartan subalgebra $\mathfrak{h}$ and
\bea\label{ineqR}
\mbox{rank}\,\mathcal{R}\ge \mbox{dim}\,\mathfrak{g}-\mbox{rank}\,\mathfrak{g}\,.
\eea
The standard solution of the classical Yang--Baxter equation \cite{BosonicYangBaxter} corresponds to the case where the last inequality saturates, so the kernel of ${\mathcal{R}_b}^a$ coincides with the Cartan subalgebra:
\bea\label{CartanR}
\mathcal{R}H_i=0\quad\mbox{for all}\quad H_i\in \mathfrak{h}.
\eea 
Looking at an arbitrary $X=H$ from this subalgebra, and representing this generator as an operator ${\hat H}$ acting in the adjoint representation, we can rewrite (\ref{mYBeq}) as
\bea\label{RHRCartan}
-\mathcal{R}{\hat H}\mathcal{R}Y={\hat H}Y.
\eea
If $Y$ is an eigenvector of $\mathcal{R}$ with an eigenvalue $\la_Y$, then ${\hat H}Y$ is an eigenvector with an eigenvalue $-\frac{1}{\la_Y}$ for any ${\hat H}$. 

To proceed, we expand the eigenvector $Y$ in the Weyl--Cartan basis,
\bea\label{YexpandCW}
Y=\sum c_k |\alpha^{(k)}\rangle,
\eea
where each $|\alpha^{(k)}\rangle$ is an eigenvector of all Cartan generators\footnote{Equation (\ref{CartWeyl}) gives a more explicit expression, but it is not needed here.}. Focusing on a particular Cartan generator 
${\hat H}_i$, we conclude that $[{\hat H}_i]^NY$ is an eigenvector of $\mathcal{R}$, which is dominated by $ |\alpha^{(k)}\rangle$ with the largest eigenvalue of ${\hat H}_i$. Removing this vector and repeating the argument for the second largest eigenvalue and so on, one can demonstrate that all 
$|\alpha^{(k)}\rangle$ are eigenvectors of $\mathcal{R}$. In other words, we have shown that matrix $\mathcal{R}$ must be diagonal in the Cartan--Weyl basis. 

Let us now specify the  Cartan--Weyl basis in more detail. Any semisimple Lie algebra admits a decomposition into the Cartan generators $H_i$ and ladder operators $E_\alpha$ so that the full commutation relations have the form
\bea\label{CartWeyl}
[H_i,H_j]=0,\quad [H_i,E_\alpha]=\alpha_i E_\alpha,\quad [E_\alpha,E_\beta]=e_{\alpha,\beta}E_{\alpha+\beta},\quad [E_\alpha,E_{-\alpha}]=\sum_i {\tilde\alpha}^i H_i\,.
\eea
In the expansion (\ref{YexpandCW}) the generator $E_\alpha$ was denoted as $|\alpha^{(k)}\rangle$. 
By an appropriate rescaling of the ladder operators one can go to a more restrictive Chevalley basis, but such specification will not play any role in our discussion. As we have demonstrated, relation (\ref{CartanR}) implies that the R--matrix must be diagonal in the basis (\ref{CartWeyl}), this leads to the explicit form of the Yang--Baxter operator:
\bea
\mathcal{R}H_i=0,\quad \mathcal{R}E_\alpha=\la_\alpha E_\alpha
\eea
Substitution into (\ref{RHRCartan}) leads to $\la_\alpha=\pm i$, and application of the Yang--Baxter equation (\ref{mYBeq}) to $(X,Y)=(E_\alpha,E_\beta)$ gives a constraint on the eigenvalues
\bea
\la_\alpha\la_\beta-\la_{\alpha+\beta}(\la_\alpha+\la_{\beta})=1.
\eea
In particular, $\la_\alpha\la_{-\alpha}=1$, so the Yang--Baxter operator becomes:
\bea\label{CanonYBsln}
\mathcal{R} H_i=0,\quad \mathcal{R} E_\alpha=-i E_\alpha,\quad \mathcal{R} E_{-\alpha}=i E_{-\alpha},
\eea
where $\alpha$ are positive roots. This construction is known as the canonical R--matrix, and we have derived it from (\ref{CartanR}), which in turn follows from the assumption that the inequality (\ref{ineqR}) saturates. 

\bigskip

The canonical R--matrix \eqref{CanonYBsln} can be easily generalized by modifying the first relation in (\ref{CanonYBsln}), and such extension will play an important role in the analysis presented in the rest of this section. Specifically, it is clear that equation (\ref{mYBeq}) is solved by
\bea\label{CanonYBslnG}
\mathcal{R}H_i={R_i}^j H_j,\quad \mathcal{R}E_\alpha=-i E_\alpha,\quad 
\mathcal{R}E_{-\alpha}=i E_{-\alpha}
\eea
for an {\it arbitrary} matrix ${R_i}^j$. In other words, the R--matrix can be modified in the Cartan subalgebra\footnote{A similar construction has been discussed in the mathematical literature \cite{Skprypnik}.}. Notice that for the deformation (\ref{CanonYBsln}) the inequality (\ref{ineqR}) is replaced by
\bea
\mbox{rank}\,\mathcal{R}= \mbox{dim}\,\mathfrak{g}-\mbox{rank}\,\mathfrak{g}+\mbox{rank}\,{R}\,.
\eea
For future reference we also give the real form of (\ref{CanonYBslnG}):
\bea
&&B_\alpha=\frac{i}{\sqrt{2}}(E_\alpha+E_{-\alpha}),\quad C_\alpha=\frac{1}{\sqrt{2}}(E_\alpha-E_{-\alpha}),\nn
\label{RoperCanon}
&&\mathcal{R}H_i={R_i}^j H_j,\quad \mathcal{R}B_\alpha=C_\alpha,\quad 
\mathcal{R}C_\alpha=-B_\alpha,
\eea
The undeformed version of this solution (i.e., the one with $R=0$) has been widely discussed in the literature \cite{BosonicYangBaxter, Skprypnik}, and the general form of (\ref{RoperCanon}) will be used later in this section.

While (\ref{CanonYBsln}) was the most general solution with saturated inequality (\ref{ineqR}), the construction (\ref{CanonYBslnG}) is just one possible option for non--saturating (\ref{ineqR}), and later we will present explicit examples of R--matrices which do not fit into (\ref{CanonYBslnG}). However, we will now demonstrate that any solution that can be obtained as a continuous perturbation of (\ref{CanonYBsln}) must have the form (\ref{CanonYBslnG}). 

\bigskip

Let us start with the canonical solution (\ref{CanonYBsln}), which will be called $\mathcal{R}_0$, and perturb it  by $\eps \mathcal{R}_1$ with a small parameter $\eps$. Applying (\ref{mYBeq}) to two elements of the Cartan subalgebra ($(X,Y)\in {\mathfrak{h}}$) and expanding the result to the first order in $\eps$, we find a system of linear constraints on $\mathcal{R}_1$:
\bea\label{ConstrR1}
-\mathcal{R}_0([\mathcal{R}_1X,Y]+[X,\mathcal{R}_1Y])=0,\quad X,Y\in \mathfrak{h},
\eea
Clearly, our ansatz (\ref{CanonYBslnG}) solves these constraints with
\bea
\mathcal{R}_1H_i={R_i}^j H_j,\quad \mathcal{R}_1E_\alpha=0,\quad 
\mathcal{R}_1E_{-\alpha}=0,\nonumber
\eea 
and since equations (\ref{ConstrR1}) are linear in $\mathcal{R}_1$, one can always subtract an appropriate solution (\ref{CanonYBslnG}) to ensure that $\mathcal{R}_1X$ has a trivial projection on the Cartan subalgebra. In other words, without the loss of generality, we can write
\bea\label{R1onX}
\mathcal{R}_1X=\sum_\alpha c_X(\alpha)E_\alpha\,,
\eea
where sum is extended over all roots of the Lie algebra, and $c_X(\alpha)$ are some numerical coefficients.
Substitution into (\ref{ConstrR1}) gives
\bea\label{ConstrR1Prime}
-\sum_{\alpha}[-c_X(\alpha)Y(\alpha)+c_Y(\alpha)X(\alpha)\Big]\Big[\mathcal{R}_0 E_\alpha\Big]=0,
\eea
where coefficients $X(\alpha)$ are defined using the commutation relations (\ref{CartWeyl}):
\bea
[X,E_\alpha]=\Big[\sum_i x^i H_i,E_{\alpha}\Big]=E_\alpha \sum_i x^i\alpha_i\quad
\Rightarrow\quad [X,E_\alpha]\equiv X(\alpha)E_\alpha.
\eea
Since the roots $E_\alpha$ are eigenvectors of $\mathcal{R}_0$ (recall (\ref{CanonYBsln})), and they are linearly independent, equation (\ref{ConstrR1Prime}) implies that\footnote{For every root $\alpha$ we can always start with $Y\in {\mathfrak{g}}$, such that $Y(\alpha)\ne 0$, so the right hand side of (\ref{ConstrR1Prime2}) is well-defined.}
\bea\label{ConstrR1Prime2}
c_X(\alpha)=X(\alpha)\frac{c_Y(\alpha)}{Y(\alpha)}\equiv c(\alpha)X(\alpha)\,.
\eea
Substitution into (\ref{R1onX}) leads to
\bea\label{ConstrR1Prime4}
\mathcal{R}_1X=\sum_\alpha X(\alpha) c(\alpha) E_\alpha\,,
\eea
where $c(\alpha)$ depends on the root, but not on the element $X$ of the Cartan subalgebra. To complete the argument, we define
\bea\label{ConstrR1Prime3}
{\tilde X}\equiv X-\eps \sum_\alpha X(\alpha) c(\alpha) \left[\frac{E_\alpha}{\mathcal{R}_0 E_\alpha}\right]E_\alpha\,.
\eea
Notice that relations (\ref{CanonYBsln}) for $\mathcal{R}_0$ imply that expressions in the square brackets are $c$--numbers equal to $\pm i$. Using (\ref{CanonYBsln}), we conclude that 
\bea\label{ConstrR1Prime5}
(\mathcal{R}_0+\eps \mathcal{R}_1){\tilde X}=O(\eps^2),
\eea
so in the leading order in $\eps$ operator $\mathcal{R}$ has the same number of zero modes as $\mathcal{R}_0$, so the solution is still given by (\ref{CanonYBsln}), but the Cartan subalgebra is rotated 
by (\ref{ConstrR1Prime3}). To simplify the discussion we started with equation (\ref{R1onX}) by subtracting the part of $\mathcal{R}_1$ that acts on the Cartan subalgebra, and in general equations (\ref{ConstrR1Prime4}) and (\ref{ConstrR1Prime5}) are replaced by
\bea\label{ConstrR1Prime6}
&&\mathcal{R}_1X=R X+\sum_\alpha X(\alpha) c(\alpha) E_\alpha\,,\nn
&&(\mathcal{R}_0+\eps \mathcal{R}_1){\tilde X}=\eps R {\tilde X}+ O(\eps^2),
\eea
while equation (\ref{ConstrR1Prime3}) remains the same. Here $R$ is an operator mapping the Cartan subalgebra on itself, so equation (\ref{ConstrR1Prime6}) is a perturbative expansion of (\ref{CanonYBslnG}).  

\bigskip

To summarize, we have demonstrated that the most general solution of the mCYB equation (\ref{mYBeq}) with $\mbox{rank}\,\mathcal{R}= \mbox{dim}\,\mathfrak{g}-\mbox{rank}\,\mathfrak{g}$ is given by (\ref{CanonYBsln}), and its most general perturbation fits the ansatz (\ref{CanonYBslnG}). It would be interesting to find the most general solution of the mCYB equation without relying on perturbative argument, but such investigation is beyond the scope of this article.

\bigskip

So far we have focused on the Yang--Baxter equation (\ref{mYBeq}) and have ignored the coset constraint 
(\ref{YBcosetConstr1}). This leads to the expression (\ref{CanonYBslnG}), which is not sensitive to the choice of the coset, but condition (\ref{YBcosetConstr1}) projects out some solutions. If fact, as we will see in subsection \ref{SubsAdS5R}, in the case of the SO(6)/SO(5) coset the constraint  (\ref{YBcosetConstr1}) eliminates all solutions preventing the construction of the generalized $\la$--deformation for AdS$_5\times$S$^5$. Note that while the construction (\ref{CanonYBslnG}) can be applied to any 
Cartan subalgebra and all resulting R--matrices would be related by a group rotation, a specific embedding of the subgroup $F$ removes equivalence between different choices of the Cartan subalgebra. Thus the 
constraint (\ref{YBcosetConstr1}) should be imposed on the R--matrices which have the form (\ref{CanonYBslnG}) for {\it at least one} Cartan subalgebra.  Starting with one Cartan subalgebra, applying the prescription (\ref{CanonYBslnG}), and rotating the result by an arbitrary element of the group, one constructs the most general R--matrix in the class (\ref{CanonYBslnG}), which depends on $N$ parameters with 
\bea
N=\frac{r(r-1)}{2}+(d-r),\qquad r=\mbox{rank}\,\mathfrak{g},\quad d=\mbox{dim}\,\mathfrak{g}.
\eea
The constraint (\ref{YBcosetConstr1}) should be imposed in the end. 

We conclude this subsection by presenting an explicit example of the construction (\ref{CanonYBslnG}), (\ref{YBcosetConstr1}) for the simplest coset SU(2)/U(1). Since SU(2) has a one--dimensional Cartan subalgebra, the antisymmetric matrix $R_{ij}$ entering (\ref{CanonYBslnG}) must be trivial, so in the real basis the $R$--matrix has only two non--zero elements:
\bea
\mathcal{R}_{12}=-\mathcal{R}_{21}=1.
\eea
Rotation by a group element leads to a more general matrix in terms of the Euler angles
\bea\label{RmtrSU2}
\mathcal{R}=\left[\begin{array}{ccc}
0 &\cos\theta & \sin\theta\cos\phi\\
-\cos\theta & 0 & \sin\theta\sin\phi\\
-\sin\theta\cos\phi & -\sin\theta\sin\phi & 0
 \end{array} \right].
\eea
Direct calculation shows that this is the most general solution of the Yang--Baxter equation (\ref{mYBeq}).
The coset constraint \eqref{CosetConstrMtr} is satisfied trivially.

In the next few subsections we will discuss some examples of cosets arising in string theory.

\subsection{Solution for SO(3)/SO(2)}

Let us discuss the most general solutions of the modified Yang-Baxter equation for the cosets SO(3)/SO(2) and SO(2,1)/SO(1,1), which arise in the deformation of AdS$_2\times$S$^2$. Strings on this background are described by the supercoset $\mathfrak{psu(1,1|2)}$ \cite{AdS2PSU}, whose bosonic sector is represented by two $2\times 2$ matrices $\mathfrak{g_{u(2)}}$, $\mathfrak{g_{u(1,1)}}$:
\bea
\mathfrak{g_{psu(1,1)}}=\left[
\begin{array}{cc}
\mathfrak{g_{u(1,1)}}& 0\\
0& \mathfrak{g_{u(2)}}
\end{array}
\right],\quad \mathfrak{g^\dagger_{u(1,1)}} \Sigma\,\mathfrak{g_{u(1,1)}}=\Sigma, \quad 
\mathfrak{g^\dagger_{u(2)}}\mathfrak{g_{u(2)}}=I,\quad
\Sigma=\left[
\begin{array}{cc}
1&0\\
0&-1
\end{array}
 \right]\,.\nonumber
\eea
We will use the following explicit parameterization of generators\footnote{Labels 6-12 are usually reserved for the fermionic generators.}:
\bea
\mathfrak{g_{u(1,1)}}&=&
\left[\begin{array}{cc}
F_1+F_4& F_2+i F_3\\
-F_2+i F_3& -F_1+F_4
 \end{array} \right],\quad
\mathfrak{g_{u(2)}}=
\left[\begin{array}{cc}
F_{13}+F_{16}& F_{14}+iF_{15}\\
F_{14}-iF_{15}& -F_{13}+F_{16}
 \end{array} \right].
\eea
U(2) subgroup has two--dimensional Cartan subalgebra spanned by $(F_{13},F_{16})$, and the construction (\ref{CanonYBsln}) gives
\bea\label{SimpleRU2}
\mathcal{R}_{U(2)}&=&\left[
\begin{array}{cccc} 
0&0&0&a \\
0 & 0 & 1& 0\\
0& -1& 0 & 0\\
-a&0& 0& 0
\end{array} 
\right]\,.
\eea
Rotation by a general group element gives
\bea\label{RmatrU2}
\mathcal{R}_{U(2)}&=&\left[
\begin{array}{cccc} 
0& \cos\gamma\sin\theta& \sin\gamma\sin\theta &a \\
-\cos\gamma\sin\theta & 0 & \cos\theta& -a \sin\gamma\tan\theta\\
-\sin\gamma\sin\theta& -\cos\theta& 0 & a\cos\gamma\tan\theta\\
-a& a \sin\gamma\tan\theta& -a\cos\gamma\tan\theta& 0
\end{array} 
\right]\,,
\eea
and direct calculation shows that this is the most general R--matrix for U(2). Choosing the subgroup $F$ spanned by $(F_{13},F_{16})$, one can check that the constraint (\ref{YBcosetConstr1}) is satisfied.

The R--matrix for U(1,1) is obtained by rotating the counterpart of (\ref{SimpleRU2}) by an appropriate group element, and the result is
\bea\label{RmatrU11}
\mathcal{R}_{U(1,1)}&=&\left[
\begin{array}{cccc} 
0& \cos\gamma\sinh\xi& \sin\gamma\sinh\xi &a \\
-\cos\gamma\sinh\xi & 0 & \cosh\xi& a \sin\gamma\tanh\xi\\
-\sin\gamma\sinh\xi& -\cosh\xi& 0 & -a\cos\gamma\tanh\xi\\
-a& -a \sin\gamma\tanh\xi& a\cos\gamma\tanh\xi& 0
\end{array} 
\right].
\eea
While constructing the integrable deformations of strings on AdS$_2\times$S$^2$, one can obtain the fields for U(1,1)/U(1) by analytic continuation of the result for U(2)/U(1). This is slightly easier than performing a separate calculations using (\ref{RmatrU11}), but the answers are the same.

\subsection{Solution for SO(4)/SO(3)}

Next, we consider the coset
\bea
\frac{SO(4)}{SO(3)}=\frac{SU(2)_L\times SU(2)_R}{SU(2)_{diag}}.
\eea
This coset, along with its counterpart $SO(2,2)/SO(1,1)$, arises in description of strings on AdS$_3\times$S$^3$.

To simplify the evaluation of the R--matrix we pick the following generators of SU(2)$\times$SU(2)
\bea\label{GeneratorsSO4}
T^{[SU(2)]^2}=\{T^L, T^R \},\quad 
T^L_i=\left[
\begin{array}{cc}
\sigma_i& 0\\
0& 0
\end{array}
 \right],\quad 
 T^R_i=\left[
\begin{array}{cc}
0& 0\\
0& \sigma_i
\end{array}
 \right],
\eea
where $\sigma_i$ are the Pauli matrices. The subgroup SU(2)$_{diag}$ is generated by
\bea
T_i^{diag}=\frac{1}{2}\left[\begin{array}{cc}
\sigma_i &0\\ 0&\sigma_i
\end{array}\right].
\eea
Starting with  the most general antisymmetric $R$ matrix 
\bea\label{RmtrSO4ansatz}
\mathcal{R}=\left[\begin{array}{cc}
A&B\\
-B^T&C
\end{array}\right]
\eea
and performing an SU(2)$_{diag}$ rotation,
we can put the antisymmetric matrix $A$ in the form
\bea
A= \left[\begin{array}{ccc}
0&0&0\\
0&0&1\\
0&-1&0
\end{array}\right]
\eea
An additional rotation in the 2--3 plane can be used to set $B_{31}=0$. 

Direct substitution of \eqref{RmtrSO4ansatz} into the modified Yang-Baxter equation \eqref{YBmatrix} and the coset constraint \eqref{CosetConstrMtr} leads to three families of the R matrices and one special solution $\mathcal{R}_4$:

\bea\label{RmatricesSO4}
\mathcal{R}_1&=\left[\begin{array}{cccccc}
0& 0& 0& a& 0& 0\\
0& 0& 1& 0& 0& 0\\
0& -1& 0& 0& 0& 0\\
-a& 0& 0& 0& 0& 0\\
0& 0& 0& 0& 0& -1\\
0& 0& 0& 0& 1& 0
 \end{array}\right],\quad&
\mathcal{R}_2=\left[\begin{array}{cccccc}
0& 0& 0& i& b& -ib\\
0& 0& 1& 0& ic& c\\
0& -1& 0& 0& c& -ic\\
-i& 0& 0& 0& b& -ib\\
-b& -ic& -c& -b& 0& -1\\
ib& -c& ic& ib& 1& 0
 \end{array}\right]\nn
\mathcal{R}_3&=\left[\begin{array}{cccccc}
0& 0& 0& -i& b& ib\\
0& 0& 1& 0& 0& 0\\
0& -1& 0& 0& 0& 0\\
i& 0& 0& 0& b& ib\\
-b& 0& 0& -b& 0& -1\\
-ib& 0& 0& -ib& 1& 0
 \end{array}\right],\quad&
\mathcal{R}_4=\left[\begin{array}{cccccc}
0& 0& 0& i& 0 & 0\\
0& 0& 1& 0& i& 1\\
0& -1& 0& 0& -1& i\\
-i& 0& 0& 0& 0& 0\\
0& -i& 1& 0& 0& 1\\
0& -1& -i& 0& -1& 0
 \end{array}\right],
\eea
As expected from the general analysis of subsection \ref{SubsRgen}, only $\mathcal{R}_1$, which fits the ansatz (\ref{CanonYBslnG}), can be continuously connected to the canonical solution (\ref{CanonYBsln}). All other matrices are complex, and they cannot be transformed into $\mathcal{R}_1$ or into each other by any 
action of SU(2)$\times$SU(2) (recall that $g\in$ SU(2)$\times$SU(2) acts as a rotation $\mathcal{R}\rightarrow g \mathcal{R} g^{-1}$). Since matrices $\mathcal{R}_{2,3,4}$ are complex, they are not acting in a proper real section of the SU(2)$\times$SU(2) algebra, so they will not play any role in our construction. Interestingly, the generalized canonical solution (\ref{CanonYBslnG}) exhausts all real R matrices. While this result was proven in subsection \ref{SubsRgen} using perturbative techniques, the current example suggests that it might hold in general. On the other hand, example (\ref{RmatricesSO4}) illustrates that in complexified algebras solution (\ref{CanonYBslnG}) is not unique beyond perturbation theory. It would be interesting to study the counterparts of $\mathcal{R}_{2,3,4}$ for other complexified algebras.

\subsection{Absence of solution for SO(6)/SO(5)}
\label{SubsAdS5R}

Finally let us apply the construction (\ref{CanonYBslnG}) to the coset 
\bea
\frac{SO(6)}{SO(5)}\,,
\eea
which arises in description of strings on AdS$_5\times$S$^5$.

The generators of $SO(6)$ are defined as 
\bea
(T_{mn})_{ab}=\delta_{ma} \delta_{nb}-\delta_{mb}\delta_{na}, \quad m,n,a,b=1,...,6,
\eea
the Cartan subgroup is three--dimensional, and it can be represented by 
\bea
H=\{T_{23}, T_{45}, T_{61}\}.
\eea
The standard diagonalization procedure leads to twelve roots:
\bea
\alpha_\alpha=\{(0,a,b), (a, 0, b), (a,b, 0)\},\qquad a,b=\pm 1.
\eea
A root will be considered positive if the first non-zero entry is positive, and for such roots prescription 
(\ref{CanonYBslnG}) gives $\mathcal{R}E_\alpha=-i E_\alpha$. For negative roots we have $\mathcal{R}E_{-\alpha}=i E_{-\alpha}$. Since $SO(6)$ has rank three, the antisymmetric matrix ${R_{ij}}$ appearing in (\ref{CanonYBslnG}) has only one non--zero element.

Next we should specify the subgroup and check the coset constraint \eqref{CosetConstrMtr}. Instead of choosing a particular subgroup, we parametrize the entire family of SO(5) embeddings, which are in the one--to--one correspondence with the unit vectors in $\mathbb{R}^6$. In the simplest case of the unit vector with only one nontrivial component $v_1=1$, the coset generators are given by 
\bea
(T^{(cos,0)}_i)_{ab}=\delta_{ia} \delta_{1b}-\delta_{ib}\delta_{1a},\quad i=2,...,6,
\eea
and in general we find
\bea\label{CosetRotate}
T^{(cos)}=(g^{SO(6)})^{-1}T^{(cos,0)}g^{SO(6)}
\eea
The SO(6) group element is parameterized in terms of the Euler angles as \cite{Fradkin}
\bea
g^{SO(n)}=\prod_{i=1}^n \prod_{j=i}^1 g_j(\theta_j^i),\quad g_k(x)=\exp\left[ x T_{n+1-k,n+1-(k+1)} \right].
\eea
and the independent choices of the cosets (\ref{CosetRotate}) correspond to $\theta_{i,5}$.
Plugging the extended canonical R--matrix \eqref{CanonYBslnG} into the coset constraint \eqref{CosetConstrMtr} we find that there are no solutions, which means that the coset SO(6)/SO(5) does not satisfy the coset constraint, and it is impossible to construct the generalized $\lambda$ deformation of AdS$_5\times$S$^5$.

\subsection{Graded Yang-Baxter equation}
\label{SecGraded}

Although in this article we are focusing on deformations of bosonic cosets, in the future it might be interesting to extend the generalized lambda deformation to supercosets describing string theories on AdS$_p\times$S$^p$ \cite{AdS2PSU,AdS3PSU,AdS5PSU}. For the ordinary lambda deformation this has been done in \cite{HMS}, but the generalized deformation is more involved. However, preliminary analysis indicates that an extension to supercoset would involve the graded Yang-Baxter equation, and in this subsection we will briefly discuss its properties and some solutions.

To define the Yang-Baxter equation on superalgebras and supercosets, one replaces the commutators in \eqref{modYBeq} by the graded commutators
\bea\label{GradedModYBeq}
[\mathcal{R} X,\mathcal{R} Y\}-\mathcal{R}([\mathcal{R} X,Y\}+[X,\mathcal{R} Y\})=-c^2 [X,Y\},\quad A,B\in \mathfrak{g},\quad c\in\mathbb{C}.
\eea
To define the graded commutator we start with supermatrices $X,Y$ written in the block form
\bea\label{supermtrs}
X=\left[
\begin{array}{c|c}
A& B\\
\hline
C&D
\end{array}
 \right],\quad 
 Y=\left[
\begin{array}{c|c}
E& F\\
\hline
G&H
\end{array}
 \right],
\eea
where the blocks in the left upper and right bottom corners are called even (bosonic), and the blocks in the right upper and left bottom corners - odd (fermionic). If terms of supermatrices \eqref{supermtrs} the graded commutator is \cite{Beisert}
\bea
[X,Y\}=
\left[
\begin{array}{c|c}
AE+BG-EA+FC& AF+BH-EB-FD\\
\hline
CE+DG-GA-HC& CF+DH+GB-HD
\end{array}
\right].
\eea
The generalized canonical R--matrix for the supercoset can be constructed by a simple extension of (\ref{CanonYBslnG}). After choosing bosonic Cartan subalgebras for blocks $A$ and $B$ in (\ref{supermtrs}), we find the roots and the counterparts of the ladder operators $E_\alpha$ in (\ref{CartWeyl}), 
\bea\label{GradedCartWeyl}
[H_i,E_\alpha\}=\alpha_i E_\alpha,
\eea
but now some of $E_\alpha$ are fermionic. Direct calculation shows that the R--matrix
\bea\label{CanonYBsuMatr}
\mathcal{R}H_i={R_i}^j H_j,\quad \mathcal{R}E_\alpha=-i E_\alpha,\quad 
\mathcal{R}E_{-\alpha}=i E_{-\alpha}
\eea
solves the graded Yang-Baxter equation (\ref{GradedModYBeq}). Let us present an explicit solution for the 
 superalgebra $\mathfrak{psu(1,1|2)}$, which arises in description of strings on AdS$_2\times$S$^2$ \cite{AdS2PSU}. 

\bigskip 

The superalgebra $\mathfrak{psu(1,1|2)}$ is defined in terms of the $4\times 4$ supermatrices
\bea
\mathcal{M}=\left[
\begin{array}{c|c}
A& B\\
\hline
C&D
\end{array}
 \right]
\eea
subject to constraint
\bea
\left[\begin{array}{cc}
A&B\\C&D
\end{array}\right]=\left[\begin{array}{cc}
\Sigma A^\dagger \Sigma^{-1}&-i\Sigma C^\dagger\\-iB^\dagger \Sigma^{-1}&D^\dagger
\end{array}\right]\,,\qquad \Sigma=\mbox{diag}(1,-1).
\eea
Parameterizing such matrix as
\bea
\mathcal{M}=
\left[
\begin{array}{cc|cc}
F_1+F_4& F_2+i F_3& F_5+i F_6& F_7+i F_8\\
-F_2+i F_3& -F_1+F_4& F_9+i F_{10}& F_{11}+i F_{12}\\
\hline
-i F_5-F_6& i F_9+F_{10}& F_{13}+F_{16}& F_{14}+i F_{15}\\
-i F_7-F_8& iF_{11}+F_{12}& F_{14}-iF_{15}& -F_{13}+F_{16}
\end{array}
\right]
\eea
and choosing the canonical solution (\ref{CanonYBsuMatr}) with $R=0$, we find $6\times 2$ nonzero elements
\bea
\mathcal{R}_{23}=1,\quad \mathcal{R}_{14,15}=1,\quad
\mathcal{R}_{5,6}=\mathcal{R}_{7,8}=-\mathcal{R}_{9,10}=-\mathcal{R}_{11,12}=-i,\quad \mathcal{R}_{ab}=-\mathcal{R}_{ba}\,.
\eea
In the alternative parametrization of the $\mathfrak{psu(1,1|2)}$ matrix in terms of the holomorphic variables, which is often used in the literature \cite{HT},
\bea\label{HolomSuMatr}
\mathcal{M}=
\left[
\begin{array}{cc|cc}
F_1+F_4& F_2+i F_3& i F_8& F_5\\
-F_2+i F_3& -F_1+F_4& i F_{6}& F_{7}\\
\hline
i F_9& -i F_{11}& F_{13}+F_{16}& F_{14}+i F_{15}\\
F_{10}& -F_{12}& F_{14}-iF_{15}& -F_{13}+F_{16}
\end{array}
\right].
\eea
the R--matrix is
\bea\label{SuYangBaxt}
\mathcal{R}_{23}=1,\quad \mathcal{R}_{14,15}=1,\quad \mathcal{R}_{9,8}=\mathcal{R}_{5,10}=-\mathcal{R}_{11,6}=-\mathcal{R}_{7,12}=\frac{i}{2},\quad
\mathcal{R}_{ab}=-\mathcal{R}_{ba}\,.
\eea
Supercoset (\ref{HolomSuMatr}) has been used to construct the standard $\la$--deformation of strings on AdS$_2\times$S$^2$, and the generalized $\la$--deformation would be based on the solution  (\ref{SuYangBaxt}) of the modified Classical Yang--Baxter equation (\ref{GradedModYBeq}). However, before constructing such solutions one should prove that the resulting deformed supercoset leads to integrable theories, as was done for the standard $\la$ deformation in \cite{HMS}, and such analysis is beyond the scope of this paper. In the remaining part of this article we will focus on bosonic cosets.

\section{SUGRA embeddings of the generalized $\lambda$--deformations}\label{SUGRAemb}
\renewcommand{\theequation}{4.\arabic{equation}}
\setcounter{equation}{0}

The general construction reviewed in section \ref{SectReview} gives the bosonic part of the string action  (\ref{gWZW}), (\ref{GenLambdaAction}) for the integrable $\la$--deformation, and in this section we will extract metric and the dilaton from these expressions. After introducing the general procedure in subsection 
\ref{SubsGenMetr}, we use it to derive the deformations of AdS$_2\times$S$^2$ and AdS$_3\times$S$^3$ in subsections \ref{SubsAdS2} and \ref{SubsAdS3}. As in the case of integrable deformations encountered earlier \cite{OtherEta,ST,SfetsosAdS5,HT,BTW, ChLLambda}, the Ramond--Ramond fluxes are recovered from solving the equations of motion of supergravity rather than from the fermionic part of the sigma model\footnote{It has been shown in \cite{BTW} that the extraction of the RR fluxes from the fermionic part of the sigma model is notoriously complicated.}.

\subsection{General construction}
\label{SubsGenMetr}

We begin with constructing the metric and the dilaton for deformations of arbitrary cosets $G/F$. To do so, we need three ingredients from section \ref{SectReview}:  the matrix $D_{ab}$, the left--invariant form $L$ parameterizing the coset, and the matrix $\hat{\lambda}^{-1}$ specifying the deformation. These ingredients are given by (\ref{lambdaMtrDef}) and (\ref{EmtrCosetA})\footnote{Most results of this subsection would apply to any matrix ${E}_G$, not only the one given in by (\ref{SmryDmatr}).}:
\bea\label{SmryDmatr}
&&D_{ab}=\mbox{Tr}(T_a g T_b g^{-1}), \quad
L_a
=i\mbox{Tr}(T_ag^{-1}d g),\nn
&&\hat{\lambda}^{-1}=(I-P){E}_G(I-P)+I,\quad
{E}_G=\frac{1}{\tilde{t}}(I-\tilde{\eta}\mathcal{R})^{-1}\,.
\eea
Here $P$ is the projector on the subgroup $F$, $\mathcal{R}$ is a solution of the modified Classical Yang--Baxter equation (\ref{YBmatrix}) satisfying the constraint (\ref{YBcosetConstr}), and $({\tilde t},{\tilde\eta})$ are free parameters. The authors of \cite{GenLambda} introduced two convenient parameters $(\la,\zeta)$ instead of $({\tilde t},{\tilde\eta})$,
\bea\label{EmtrYB}
\tilde{t}=\frac{\lambda}{(1-\lambda)},\quad \tilde{\eta}=-\frac{\zeta (2 \tilde{t}+1)}{2\tilde{t}},
\eea
and to compare with the existing literature, our final solution will be expressed in terms of $(\la,\zeta)$. Note, however, that the deformation depends on $(\la,\zeta)$ {\it and} all free parameters appearing in the R--matrix, so the generalized $\la$--deformation can produce {\it very large families} of integrable string theories.

The metric can be extracted from the symmetric part of the action (\ref{gWZW}), (\ref{GenLambdaAction})\footnote{Here we expressed everything in terms of $L$ using $R=DL$ and the orthogonality relation $D^TD=1$.}:
\bea\label{MetricA}
ds^2=\frac{k}{4\pi}L^T[I-\mathfrak{D}D-(\mathfrak{D}D)^T]L,\quad 
\mathfrak{D}\equiv[D-\hat{\lambda}^{-1}]^{-1}\,.
\eea
To rewrite this in terms of frames, we perform some algebraic manipulations which lead to
\bea\label{MetrPreFrame}
ds^2=\frac{k}{4\pi}L^T(\hat{\lambda}^{-1}-D)^{-1}[\hat{\lambda}^{-1}\hat{\lambda}^{-T}-I](\hat{\lambda}^{-1}-D)^{-T}L.
\eea
In the case of the isotropic deformation, where $\hat{\lambda}$ is proportional to the identity matrix, the expression in the square brackets is a constant, so the frames are given by
\bea
e=\sqrt{\frac{k(\lambda^{-2}-1)}{4\pi}}[\hat{\lambda}^{-1}-D]^{-T}L.
\eea
In general we begin with diagonalizing the symmetric matrix $\hat{\lambda}^{-1}\hat{\lambda}^{-T}$ using an orthogonal transformation $A$:
\bea\label{DefAmatr}
\hat{\lambda}^{-1}\hat{\lambda}^{-T}=A\Lambda^{-2}A^T,\quad AA^T=I,
\eea
then the metric (\ref{MetrPreFrame}) can be recovered from the frames
\bea\label{frames}
e=\sqrt{\frac{k}{4\pi}}\sqrt{\Lambda^{-2}-I}A^T[\hat{\lambda}^{-1}-D]^{-T}L.
\eea
Note that a general $n\times n$ matrix $\hat{\lambda}^{-1}$ can be parameterized in terms of a diagonal matrix $\Lambda$ and two orthogonal matrices $A$, $B$:
\bea\label{DgnlzLam}
\hat{\lambda}^{-1}=A{\Lambda}^{-1}B,\qquad AA^T=I,\quad BB^T=I\,,
\eea
and for computational purposes we will use a slightly different but equivalent expression for the frames:
\bea\label{simpleFrames}
e
&=&\sqrt{\frac{k}{4\pi}}\sqrt{I-\Lambda^2}[(I-D^T \hat{\lambda}^T) B^{-1} ]^{-1}L.
\eea

The dilaton is defined analogously to the regular $\lambda$-deformation \cite{ST}
\bea\label{Dilaton}
e^{-2\Phi}=e^{-2\Phi_0}\mbox{det}[\hat{\lambda}^{-1}-D].
\eea
One can also extract the Kalb--Ramond field by taking an antisymmetric part of the action (\ref{GenLambdaAction}), but such $B$ field vanishes in all our examples, so it will not be discussed further.

\bigskip
 
Expressions  (\ref{frames}) and (\ref{Dilaton}) have some remarkable properties which follow from the structure of matrices $D$ and ${\hat\la}$. As shown in the appendix,
\begin{quote}
For any coset $G/F$ there exists a canonical gauge, where matrix $\mathfrak{D}=[D-\hat{\lambda}^{-1}]^{-1}$ has three properties:
\begin{enumerate}[(i)]
\item matrix $(I-P)\mathfrak{D}(I-P)$ has constant entries;
\item matrix $\mathfrak{D}(I-P)$ factorizes as $\mathfrak{D}(I-P)=ST$, where $S$ does not depend on the deformation, and $T$ is a constant matrix;
\item the dependences upon coordinates and constant deformation parameters factorizes in 
$[\mbox{det}\, \mathfrak{D}]$.
\end{enumerate}
\end{quote}
The canonical gauge is defined by the commutation relations (\ref{CanonGauge}), and such gauge will be imposed throughout this article. We will now demonstrate that properties (i)--(iii) lead to drastic simplifications in the frames (\ref{frames}) and in the dilaton (\ref{Dilaton}). 

\bigskip

The implication for the dilaton is obvious: property (iii) ensures that the deformation parameters appear in (\ref{Dilaton}) only in a constant prefactor, and thus they can be absorbed into a shift of $\Phi_0$. For specific examples this property has been seen in \cite{SfetsosAdS5}, but the analysis presented in the appendix establishes the factorization in full generality. It is worth mentioning that in the case of the ordinary $\la$--deformation (i.e., for $\zeta=0$), the metric (\ref{MetricA}) can support two integrable string theories: one is based on the coset construction, and its dilaton is given by (\ref{Dilaton}) \cite{ST,SfetsosAdS5}, while the alternative is based on super--coset, and the resulting dilaton does not factorize between the coordinates and the deformation parameters \cite{HMS,HT,BTW,ChLLambda}. It would be very interesting to find the supercoset counterpart of  (\ref{Dilaton}) for nonzero $\zeta$, but such investigation is beyond the scope of this article. 

To find the implications of the properties (ii)--(iii) for the frames, we rewrite equation  (\ref{frames}) as
\bea\label{framesCurlD}
e=-\sqrt{\frac{k}{4\pi}}\sqrt{\Lambda^{-2}-I}A^T\mathfrak{D}^{T}L.
\eea
Recalling that $P\hat{\lambda}^{-1}=\hat{\lambda}^{-1}P=P$ (see (\ref{SmryDmatr})), we conclude that 
matrices $(A,B,\Lambda)$ in (\ref{DgnlzLam}) {\it can be chosen} in such a way that\footnote{Since matrix  $\hat{\lambda}^{-1}$ has degenerate eigenvalues, relation (\ref{DefAmatr}) does not define $A$ uniquely. In addition, one has a freedom of permuting eigenvalues, and equation (\ref{ChooseABLa}) would be satisfied only for a particular ordering.} 
\bea\label{ChooseABLa}
PA=AP=P,\quad PB=BP=B,\quad \Rightarrow\quad P\Lambda=\Lambda P=P.
\eea
Introducing an explicit split between the generators of the subgroup $F$ and the coset $G/F$, one can rewrite (\ref{ChooseABLa}) more explicitly:
\bea
A=\left[\begin{array}{cc}
I&0\\
0&{\tilde A}
\end{array}\right],\quad
B=\left[\begin{array}{cc}
I&0\\
0&{\tilde B}
\end{array}\right],\quad
\Lambda=\left[\begin{array}{cc}
I&0\\
0&{\tilde \Lambda}
\end{array}\right].
\eea
Relations (\ref{ChooseABLa}) imply that 
\bea
\sqrt{\Lambda^{-2}-I}=(I-P)\sqrt{\Lambda^{-2}-I}(I-P),
\eea
then, using the property $P^T=P$, the frames (\ref{framesCurlD}) can be rewritten as
\bea
e=-\sqrt{\frac{k}{4\pi}}[I-P]\sqrt{\Lambda^{-2}-I}A^T\Big[\mathfrak{D}[I-P]\Big]^{T}L.
\eea
Application of the property (iii) leads to the final result:
\bea\label{CosetFrmFnl}
e=-\sqrt{\frac{k}{4\pi}}[I-P]\Big(\sqrt{\Lambda^{-2}-I}[TA]^T\Big)\Big(S^{T}L\Big)\,.
\eea
Equation (\ref{CosetFrmFnl}) has three distinct matrix factors. The first one ensures that frames point only along the coset directions. The second factor depends on the deformation, but not on the spacetime. The last factor gives the frames of the undeformed background, and it is not modified by the deformation. Thus application of the generalized $\la$--deformation (\ref{SmryDmatr}) simply rotates the frames by 
{\it constant} matrices. This feature has been observed for several explicit examples \cite{ST,SfetsosAdS5}, but it is proven in full generality by the analysis presented here and in the Appendix.

\subsection{Deformation of AdS$_2\times$S$^2$} 
\label{sec:supergravity_background_for_ads2}
\label{SubsAdS2}

In this subsection we embed the generalized $\lambda$-deformation of $\frac{SU(2)}{U(1)}\times \frac{SU(1,1)}{U(1)}$ into the type IIB supergravity. First we discuss the coset 
$G/F\equiv SU(2)/U(1)$ corresponding to the sphere, and the AdS part of the geometry will be obtained by an analytic continuation. 

The embedding of $F=U(1)$ into $G=SU(2)$ is unique up to an SU(2) rotation, so without loss of generality we choose the generators of $F$ and $G/F$ as 
\bea\label{S2generat}
F:\{\sigma_3\}\,,\qquad G/F:\{\sigma_1,\sigma_2 \}\,.
\eea
A general element of SU(2) can be written as 
\bea
g=e^{i(\phi_1-\phi_2)\sigma_3/2}e^{i\omega\sigma_1}e^{i(\phi_1+\phi_2)\sigma_3/2}\,,
\eea
and the gauge freedom corresponding to U(1) is fixed by setting $\phi_2=0$. As discussed in the end of subsection \ref{SubsRgen}, the R--matrix for SU(2) is unique up to a global rotations parameterized by two Euler angles (see (\ref{RmtrSU2})), but since we have already chosen the embedding of $F$ into $G$, the deformations related by global rotations may not be equivalent. Since the rotation in $(\sigma_1,\sigma_2)$ plane does not distort the embedding (\ref{S2generat}), R--matrices  (\ref{RmtrSU2}) with different angles $\phi$ lead to equivalent deformations, but dependence on the parameter $\theta$ is nontrivial. Thus the most general deformation of the SU(2)/U(1) coset is parameterized by the R--matrix
\bea\label{RmatrAdS2}
\mathcal{R}=\left[ 
\begin{array}{ccc}
0& \cos\theta& \sin\theta\\
-\cos\theta & 0 & 0\\
-\sin\theta & 0 & 0
\end{array}
\right]\, .
\eea
We begin with discussion of the simplest deformation with $\theta=0$, and we will comment on the general case in the end of this subsection. The deformation matrix $\hat{\lambda}$ is evaluated using equations \eqref{SmryDmatr}, \eqref{EmtrYB} and the projector
\bea
P=\left[\begin{array}{ccc}
0& 0& 0\\
0& 0& 0\\
0& 0& 1
 \end{array}\right].
\eea
Then equation \eqref{simpleFrames} gives the explicit expression for the frames, and to simplify them, we introduce new coordinates $(p,q)$ following \cite{HT}:
\bea
\omega=\arccos\sqrt{p^2+q^2},\quad \phi_1=\arccos\frac{p}{\sqrt{p^2+q^2}}.
\eea
The frames become
\bea\label{FramesS2}
e^i&=&U^{i}{}_je^j_{(0)},\quad e^1_{(0)}=\sqrt{\frac{k}{2\pi(1-p^2-q^2)}}dp,\quad e^2_{(0)}=\sqrt{\frac{k}{2\pi(1-p^2-q^2)}}dq,\\
U^i{}_j&=&\frac{1}{\sqrt{(1-\lambda^2)(4\lambda^2+(1+\lambda)^2\zeta^2)}}\left[
\begin{array}{cc}
-(1+\lambda)(\zeta^2+\lambda(2+\zeta^2)) & \zeta(1-\lambda^2)\\
-(1-\lambda^2)\zeta & -2(1-\lambda)\lambda
\end{array}
 \right],\nonumber
\eea
where $i,j=1,2$. The metric and the $SU(2)$ contribution to the dilaton (see \eqref{Dilaton}) are
\bea\label{defS2metr}
2\pi k^{-1}ds_S^2&=&\frac{(1+\lambda)^2(1+\zeta^2)dp^2+2(1-\lambda^2)\zeta dp dq+(1-\lambda)^2dq^2}{(1-p^2-q^2)(1-\lambda^2)}\,,\\
\label{lambdaS2dil}
e^{-2\Phi_S}&=&1-p^2-q^2.
\eea
The AdS$_2$ counterparts of the metric and the dilaton are found by performing the analytic continuation which has been used in the case of the regular $\la$ deformation \cite{ST},
\bea
q\to i y,\quad p\to x,\quad k\to -k\,,
\eea
and the result is
\bea\label{defAdS2metr}
2\pi k^{-1}ds_{AdS}^2&=&-\frac{(1+\lambda)^2(1+\zeta^2)dx^2+2i(1-\lambda^2)\zeta dx dy-(1-\lambda)^2dy^2}{
(1-x^2+y^2)(1-\lambda^2)}\, .\nn
e^{-2\Phi_{AdS}}&=&-(1-x^2+y^2).
\eea
Note that the dilaton is real since we are working in the domain where $1-x^2+y^2<0$. 

The Ramond--Ramond fluxes can be found by solving the equations of motion for type IIB supergravity
\bea
&&\nabla^2 e^{-2\Phi}=0,\nn
&&\d_m \left(\sqrt{-g} F^{mn} \right)=0,\nn
&&R_{mn}+2\nabla_m\nabla_n \Phi=\frac{e^{2\Phi}}{2}\left( F_{mk}F_n{}^k-\frac{1}{4}g_{mn}F_{ij}F^{ij} \right),
\eea
and the result is\footnote{For example, one can start  for the $\lambda$-deformation, which corresponds to $\zeta=0$, and develop the perturbation theory in $\zeta$.} 
\bea\label{defAdS2flux}
F^{(2)}&=&c_1[S\zeta (dxdp-i dydq)-S^{-1}dxdq]+c_2[S\zeta(idxdp+dydq)+S^{-1}dydp],\nn
S&=&\sqrt{\frac{1-\lambda^2}{4\lambda+(1+\lambda)^2\zeta^2}},\quad c_1^2+c_2^2=\frac{2k}{\pi}.
\eea

Notice that the metric (\ref{defAdS2metr}) and the flux (\ref{defAdS2flux}) are complex unless $\zeta=0$. This is a peculiar feature of the generalized lambda deformation of AdS$_2\times$S$^2$, which does no persist for AdS$_3\times$S$^3$ (the metric and the fluxed are real there). Although the metric (\ref{defAdS2metr})  can be made real by an additional continuation of $y$ ($y\rightarrow i y$), this procedure is not very appealing since even the undeformed metric ($\la=\zeta=0$) has a wrong signature (2,2) and a wrong isometry SO(3)$\times$SO(3). Moreover, the fluxes remain complex. 

To compare the geometry (\ref{defS2metr}), (\ref{defAdS2metr}) with the standard lambda deformation constructed in \cite{ST}, we rescale coordinates by a convenient quantity \cite{HT}
\bea
\kappa=\frac{1-\la}{1+\la}
\eea
This leads to the solution
\bea\label{RescaledAdS2sln}
\frac{2\pi}{k}ds^2&=&\frac{dp^2+(dq+\zeta dp)^2}{1-\k p^2-\k^{-1} q^2}-
\frac{dx^2-(dy-i\zeta dx)^2}{1-\k x^2+\k^{-1}y^2}\\
F^{(2)}&=&c_1[S\zeta (\k dxdp-i\k^{-1} dydq)-S^{-1}dxdq]+
c_2[S\zeta(i\k dxdp+\k^{-1}dydq)+S^{-1}dydp]\nn
e^{2\Phi}&=&-\frac{1}{(1-\k p^2-\k^{-1} q^2)(1-\k x^2+\k^{-1}y^2)}\,,\nonumber
\eea
which generalizes the geometry ({2.7}) of \cite{BTW}.

For the standard $\la$ deformation (i.e., for $\zeta=0$), the AdS$_2\times$S$^2$ geometry is recovered in the limit of small $\k$ \cite{HT}, and application of such limit to (\ref{RescaledAdS2sln}) leads to a very simple $\zeta$--dependence after some shifts and rescaling of coordinates. Indeed, the leading order in $\k$ is
\bea
\frac{2\pi}{k\k}ds^2&=&-\frac{dp^2+(dq+\zeta dp)^2}{q^2}-
\frac{dx^2-(dy-i\zeta dx)^2}{y^2}\\
F^{(2)}&=&\frac{c_1}{\sqrt{\k}}[-i{\tilde S}\zeta dydq-{\tilde S}^{-1}dxdq]+
\frac{c_2}{\sqrt{\k}}[{\tilde S}\zeta dydq+{\tilde S}^{-1}dydp]\nn
e^{2\Phi}&=&\frac{\k^2}{q^2y^2}\,,\quad {\tilde S}=\frac{1}{\sqrt{1+\zeta^2}}\nonumber
\eea
In the new coordinates defined as 
\bea
{\tilde x}= \frac{1}{1+\zeta^2}\left[x+\frac{i\zeta y}{{1+\zeta^2}}\right],\quad 
{\tilde p}=\frac{1}{1+\zeta^2}\left[p+\frac{\zeta q}{{1-\zeta^2}}\right],
\eea
the metric and fluxes become real, and $\zeta$ appears only in the radius of the AdS$_2\times$S$^2$ and in the overall normalization of the fluxes:
\bea\label{LimitAdS2def}
\frac{2\pi}{k\k}ds^2&=&\frac{1}{1+\zeta^2}\left[-\frac{d{\tilde p}^2+dq^2}{q^2}-
\frac{d{\tilde x}^2-dy^2}{y^2}\right],\quad e^{2\Phi}=\frac{\k^2}{q^2y^2}\,,\nn
F^{(2)}&=&\frac{1+\zeta^2}{\sqrt{\k}}[-c_1d{\tilde x}dq+c_2 dyd{\tilde p}],\quad c_1^2+c_2^2=\frac{2k}{\pi}\,.
\eea
To summarize, the generalized $\la$--deformation of AdS$_2\times$S$^2$ is given by (\ref{RescaledAdS2sln}). For generic values of $\la$ and nonzero $\zeta$ the fluxes and metric are complex, if one insists on the correct signature. In the $\la=1$ limit one finds the 
real solution (\ref{LimitAdS2def}), and apart from a very simple $\zeta$ dependence, it coincides with analytic continuation of AdS$_2\times$S$^2$  discussed in \cite{BTW}. 

\bigskip

We conclude this subsection by writing the solution corresponding to the general R--matrix (\ref{RmatrAdS2}). To simplify the result, it is convenient to redefine the deformation parameters as
\bea
a&=&\frac{4\lambda^2+(1-\cos^2\theta(1-\lambda))(1+\lambda)^2\zeta^2}{4\lambda+(1-\cos^2\theta(1-\lambda))(1+\lambda)^2\zeta^2},\quad 
b=-\frac{2\cos\theta\lambda(1-\lambda^2)\zeta}{4\lambda+(1-\cos^2\theta(1-\lambda))(1+\lambda)^2\zeta^2},\nn
c&=&\frac{\lambda(4\lambda+(1+\lambda)^2\zeta^2)}{4\lambda+(1-\cos^2\theta(1-\lambda))(1+\lambda)^2\zeta^2},
\eea
This brings matrix $\hat{\lambda}$ into a simple form, 
\bea
\hat{\lambda}=\left[ 
\begin{array}{ccc}
a& -b& 0\\
b& c& 0\\
0&0& 1
\end{array}
\right].
\eea
and the deformed metric becomes
\bea
2\pi k^{-1} ds_S^2&=&\frac{(1+b^2+ac+a+c)dp^2+4b dp dq+(1+b^2+ac-a-c)dq^2}{(1-b^2-ac-a+c)(1-p^2-q^2)}.
\eea
The expressions for the fluxes are not very illuminating.

\subsection{Deformation of AdS$_3\times$S$^3$} 
\label{sec:supergravity_background_for_ads3}
\label{SubsAdS3}

In this subsection we construct SUGRA embedding of the generalized lambda-deformation based on the  coset
\bea
\frac{SU(2)\times SU(2)}{SU(2)_{diag}} \times \frac{SU(1,1)\times SU(1,1)}{SU(1,1)_{diag}}.
\eea
The element of the first coset can be conveniently parameterized as 
\bea\label{CosetElementS3}
g=\left(\begin{array}{cc}
g_l&0\\ 0&g_r
\end{array}\right),\qquad g^\dagger g=I
\eea
with
\bea\label{GandGprime}
g_l=\left[\begin{array}{cc}
\alpha_0+i\alpha_3&\alpha_2+i\alpha_1\\
-\alpha_2+i\alpha_1&\alpha_0-i\alpha_3
\end{array}\right],\qquad
g_r=\left[\begin{array}{cc}
\beta_0+i\beta_3&\beta_2+i\beta_1\\
-\beta_2+i\beta_1&\beta_0-i\beta_3
\end{array}\right]\,.
\eea
The variables $\alpha_k, \beta_k$ introduced in \cite{ST} are subject to two constraints
\bea\label{DetConstS3}
\sum (\alpha_k)^2=1,\qquad \sum (\beta_k)^2=1.
\eea
Following \cite{ST}, we fix the gauge for $SU(2)_{diag}$ by setting
\bea
\alpha_2=\alpha_3=\beta_3=0,
\eea
and solve the constraints \eqref{DetConstS3} by introducing a convenient variable $\gamma$:
\bea\label{OurVariablesdS3}
 \beta_1\equiv \frac{\gamma}{\sqrt{1-\alpha_0^2}},\qquad\alpha_1=\sqrt{1-\alpha_0^2},\quad \beta_2=\sqrt{1-\beta_0^2-\frac{\gamma^2}{{1-\alpha_0^2}}}\,.
\eea
Note that the three remaining coordinates $\alpha\equiv \alpha_0$, $\beta\equiv \beta_0$ and $\gamma$ have the following ranges:
\bea\label{Range1}
0<\alpha^2<1,\quad 0<\beta^2<1,\quad \gamma^2<(1-\alpha^2)(1-\beta^2)\,.
\eea
The generators corresponding to the subgroup and the coset are related to (\ref{GeneratorsSO4}) by a linear transformation:
\bea\label{SO4SO3gens}
F:&&T_a=\frac{1}{2}\left[\begin{array}{cc}
\sigma_a &0\\ 0&\sigma_a
\end{array}\right]=\frac{1}{2}[T^L_a+T^R_a],\quad a=1,2,3;\nn
G/F:&&T_\alpha=\frac{1}{2}\left[\begin{array}{cc}
\sigma_{\alpha-3} &0\\ 0&-\sigma_{\alpha-3}
\end{array}\right]=\frac{1}{2}[T^L_{\alpha-3}+T^R_{\alpha-3}],\quad \alpha=4,5,6.
\eea
In this basis the matrix $\mathcal{R}_1$ from (\ref{RmatricesSO4}) becomes
\bea
\mathcal{R}=\left[\begin{array}{cccccc} 
0& 0& 0& 0& -1 & 0\\
0& 0& 0& 1& 0& 0\\
0& 0& 0& 0& 0& a\\
0& -1& 0& 0& 0& 0\\
1& 0& 0& 0& 0& 0\\
0& 0& -a& 0& 0& 0\\
 \end{array} \right].
\eea
The deformation matrix $\hat{\lambda}$ is obtained from \eqref{SmryDmatr}, \eqref{EmtrYB}, where the projector on the subgroup is
\bea
P=\left[
\begin{array}{cc}
I_{3\times 3}&0\\
0&0
\end{array}
\right].
\eea
Evaluation of frames using \eqref{simpleFrames} gives
\bea\label{FramesS3}
e^4_{(0)}&=&-\frac{d\alpha}{\sqrt{1-\alpha^2}},\quad
e^5_{(0)}=\left[\frac{\gamma d\alpha+(1-\alpha^2)d\beta}{\gamma'\sqrt{1-\alpha^2}}\right],\quad
e^6_{(0)}=-\frac{\beta d\alpha+\alpha d\beta-d\gamma}{\gamma'},\nn
e^4&=&c_1e^4_{(0)},\quad e^5=c_1 e^5_{(0)},\quad e^6=c_2 e^6_{(0)},\\
c_1&=&\sqrt{\frac{k}{2\pi}}\sqrt{\frac{(1+\lambda)(\zeta^2+\lambda(2+\zeta^2))}{\lambda(1-\lambda)}},\quad c_2=\sqrt{\frac{k}{2\pi}}\sqrt{\frac{\lambda(1-\lambda)}{(1+\lambda)(2\lambda+a^2\zeta^2(1+\lambda))}}.\nonumber
\eea
where we defined
\bea
\gamma'&=&\sqrt{(1-\alpha^2)(1-\beta^2)-\gamma^2}.
\eea
Interestingly, the frames (\ref{FramesS3}) depend on $\la$ and $\zeta$ only through constant prefactors, exactly as it happened for the standard $\la$--deformation \cite{ST,SfetsosAdS5}. This feature is guaranteed by the general discussion presented in subsection \ref{SubsGenMetr}. Frames (\ref{FramesS3}) exhibit one more interesting feature\footnote{We thank Ben Hoare for making this observation.}: four parameters 
$(k,\la,a,\zeta)$ appear only through two independent combinations $(c_1,c_2)$. This implies that the generalized lambda deformation describes the same set of geometies as its standard counterpart \cite{ST,SfetsosAdS5}. It would be very interesting to see whether the same feature persists for other cosets. 

\bigskip

The AdS counterpart of (\ref{FramesS3}) is obtained by performing an analytic continuation
\bea\label{AnalContAdS3}
\alpha\to \tilde{\alpha},\quad \beta\to \tilde{\beta},\quad \gamma\to \tilde{\gamma},\quad k\rightarrow -k,
\eea
and changing the the range of coordinates from (\ref{Range1})
to
\bea\label{Range2}
1<\tilde{\alpha}^2,\quad 1<\tilde{\beta}^2,\quad \tilde{\gamma}^2<(\tilde{\alpha}^2-1)(\tilde{\beta}^2-1).
\eea
Relation \eqref{Dilaton} gives the dilaton 
\bea\label{DilAdS3}
e^{-2\Phi}=e^{-2\Phi_0} \gamma'{\tilde\gamma}'\,,
\eea
and for the Ramond--Ramond fluxes, we take a simple ansatz inspired by the regular $\lambda$--deformation \cite{ST}:
\bea\label{F3ansatz}
F^{(3)}&=&C\gamma'\tilde{\gamma}'\left[e^3_{(0)}\wedge e^4_{(0)}\wedge e^5_{(0)}+e^1_{(0)}\wedge e^2_{(0)}\wedge e^6_{(0)} \right]\,.
\eea
Here $C$ is an unknown constant, which is determined by solving the equations of type IIB supergravity reduced to six dimensions:
\bea
&&\nabla^2 e^{-2\Phi}=0,\nn
&&\d_m \left(\sqrt{-g} F^{mnp} \right)=0,\nn
&&R_{mn}+2\nabla_m\nabla_n \Phi=\frac{e^{2\Phi}}{4}\left( F_{mkl}F_n{}^{kl}-\frac{1}{6}g_{mn}F_{ijk}F^{ijk} \right)\,.
\eea
The final answer is
\bea\label{FinalCAdS3}
C=\frac{k\sqrt{16\lambda^3+2(1+a^2)\lambda(1+\lambda)^3+a^2(1+\lambda)^4\zeta^4}\sqrt{\zeta^2+\lambda(2+\zeta^2)}}{4\pi(1-\lambda)\lambda\sqrt{2\lambda+a^2\zeta^2(1+\lambda)}}.
\eea
and in contrast to the deformation of $AdS_2\times S^2$, the solution (\ref{FramesS3}), (\ref{AnalContAdS3}), (\ref{F3ansatz}), (\ref{FinalCAdS3}) is real.


\section{Discussion}

In this article we have elaborated on the general procedure of constructing generalized $\la$--deformations of coset CFTs, and we have found several explicit solutions relevant for string theory. The main results of this paper can be separated into three categories. 

In section \ref{Rmatrix} we found rather general solutions of the modified classical Yang--Baxter (mCYB) equation for arbitrary cosets and supercosets, and we also constructed the {\it most general} R--matrices for the cosets arising in string theory. It would be very interesting to find the most general solutions of the mCYB for any (super)coset and to apply the results of our section \ref{SecGraded} toward generalizing the $\la$--deformation of supercosets discussed in \cite{HMS}.

The second category of our results concerns insights into the analytical structure of the generalized $\la$--deformations. In section \ref{SubsGenMetr} we demonstrated that under and {\it arbitrary} deformation of an {\it arbitrary} coset, the frames are rotated by a {\it constant} matrix and the dilaton is multiplied by a {\it constant} factor. These properties have been observed a-posteriori in several specific examples \cite{ST,SfetsosAdS5}, but our general proof allows one to drastically simplify calculations by focusing on the relevant constant matrices rather than evaluating coordinate--dependent frames.  

Finally, in sections \ref{SubsAdS2}, \ref{SubsAdS3} we constructed the generalized $\la$--deformations of AdS$_2\times$S$^2$ and AdS$_3\times$S$^3$, including the relevant Ramond--Ramond fluxes. Interestingly, while the solution corresponding to AdS$_3\times$S$^3$ is real, the deformation of AdS$_2\times$S$^2$ leads to complex metric and fluxes. It would be interesting to get a better analytical understanding of this phenomenon. In 
the AdS$_5\times$S$^5$ case we demonstrated that the construction introduced in \cite{GenLambda} does not lead to new solutions beyond the standard $\la$--deformation. 

\section*{Acknowledgments}

We thank Ben Hoare and Arkady Tseytlin for comments on the manuscript. 
OL thanks the organizers of the program ``Mathematics and Physics at the Crossroads'' at INFN -- Laboratori Nazionali di Frascati for hospitality. This work was supported by NSF grant PHY-1316184.

\appendix

\section{Properties of the matrix $D$}\label{AppDmtr}
\renewcommand{\theequation}{A.\arabic{equation}}
\setcounter{equation}{0}

In this appendix we study some properties of the matrix\footnote{For the reason which will become clear below, in this appendix we use capital letters $(A,B)$ to denote indices on the algebra $\mathfrak{g}$. This is a minor change of notation in comparison with (\ref{lambdaMtrDef}), which was more convenient in the main text.}
\bea\label{DabApp}
D_{AB}=\mbox{Tr}(T_A g T_B g^{-1}),
\eea
which plays the central role in constructing the generalized $\la$--deformation. While some empirical evidence for these properties has been accumulated from the impressive explicit calculations performed on a case--by--case basis \cite{ST,SfetsosAdS5}, to our knowledge, a general study of matrix $D_{AB}$ has not been carried out.  Using group theory, we derive several important features of this matrix which significantly simplify the construction of integrable deformations for arbitrary cosets in comparison with the explicit calculations performed in \cite{ST, SfetsosAdS5} and explain the nice `surprising relations' observed in these articles.

\bigskip

We begin with recalling the context in which matrix $D_{AB}$ arises in the $\la$--deformation of cosets. The metric is constructed using the frames (\ref{simpleFrames}), the dilaton is given by (\ref{Dilaton}), and both relations contain the expression 
\bea\label{AppEqn1}
\mathfrak{D}=[D-\hat{\lambda}^{-1}]^{-1}\,.
\eea
To construct the deformation of a coset $G/F$, one takes $g\in G/F$ and a constant matrix $\hat{\lambda}^{-1}$ given by (\ref{SmryDmatr})
\bea
\hat{\lambda}^{-1}=I+(I-P){E}_{G}(I-P)\,.
\eea
Here $P$ is a projection on a subgroup $F$, and the explicit form of matrix ${E}_{G}$, given by (\ref{SmryDmatr}), will not be important for our group theoretic discussion here. The results of this appendix can be summarized in the following statement:
\begin{quote}
For any coset $G/F$ there exists a canonical gauge (\ref{CanonGauge}), where matrix $\mathfrak{D}$ has three properties:
\begin{enumerate}[(i)]
\item matrix $(I-P)\mathfrak{D}(I-P)$ has constant entries;
\item matrix $\mathfrak{D}(I-P)$ factorizes as $\mathfrak{D}(I-P)=ST$, where $S$ does not depend on the deformation, and $T$ is a constant matrix;
\item the dependences upon coordinates and constant deformation parameters factorizes in 
$[\mbox{det}\, \mathfrak{D}]$.
\end{enumerate}
\end{quote}
By choosing the canonical gauge in sections \ref{sec:supergravity_background_for_ads2} and \ref{sec:supergravity_background_for_ads3}, we found a very simple deformation dependence in the dilatons  (\ref{lambdaS2dil}), (\ref{DilAdS3}) and frames (\ref{FramesS2}), (\ref{FramesS3}), in agreement with the general statements above.  The specific examples discussed in \cite{ST, SfetsosAdS5} provide additional illustrations of these statements.

\bigskip

We begin with specifying the convenient canonical gauge. The coset $G/F$ introduces a decomposition of the Lie algebra into a subalgebra $\mathfrak{f}$ and the remaining space $\mathfrak{l}$, and in this appendix the generators of $\mathfrak{f}$ and $\mathfrak{l}$ will be denotes using different 
labels\footnote{This decomposition shows the convenience of denoting indices in (\ref{DabApp}) by capital letters.}:
\bea
T_A\in \mathfrak{g}=\mathfrak{f}+\mathfrak{l},\qquad T_a\in \mathfrak{f}, \quad T_\alpha\in \mathfrak{l}\,.
\eea
Algebra $\mathfrak{f}$ closes under commutations, while the commutators of $T_\alpha$ are gauge--dependent, and we will choose a convenient gauge where the structure constants have only three nontrivial blocks:
\bea\label{CanonGauge}
[T_a,T_b]=\sum_c i{f_{ab}}^c T_c\,,\quad
[T_a,T_\beta]=\sum_\gamma i{f_{a\beta}}^\gamma T_\gamma\,\quad
[T_\alpha,T_\beta]=\sum_\gamma i{f_{\alpha\beta}}^c T_c\,.
\eea
In this gauge the Killing metric $\eta_{AB}\propto {f_{AM}}^N {f_{BN}}^M$ splits into two blocks $(\eta_{ab},\eta_{\alpha\beta})$ with vanishing off--diagonal elements $\eta_{a\alpha}=0$.

\bigskip

Our statement (i) reduces to coordinate independence of $\mathfrak{D}_{\alpha\beta}$, and to prove this, as well as the properties (ii) and (iii),  we begin with writing matrices $D$ and ${\hat\la}^{-1}$ in the canonical basis:
\bea\label{DfromH}
\mathfrak{D}^{-1}=D-\hat{\lambda}^{-1}=\left[\begin{array}{cc}
D_{ab}-\delta_{ab}&D_{a\beta}\\
D_{\alpha b}&D_{\alpha\beta}-H_{\alpha\beta}
\end{array}\right]\,,\quad H_{\alpha\beta}=(I+{E}_G)_{\alpha\beta}\,.
\eea
Notice that the all information about the deformation is contained in the {\it constant matrix} $H_{\alpha\beta}$, which has indices only on the coset. To proceed it is convenient to label various components of (\ref{DfromH}) by different letters:
\bea
\mathfrak{D}^{-1}\equiv \left[\begin{array}{cc}
A&B\\
C&F-H
\end{array}\right]\,.
\eea
To invert the matrix $\mathfrak{D}^{-1}$ and to compute its determinant, we introduce a triangular decomposition:\footnote{In a special case an analogous decomposition was used in \cite{SfetsosAdS5}.}
\bea\label{TriangDecomp}
\mathfrak{D}^{-1}=\left[\begin{array}{cc}
A&0\\
C&M
\end{array}\right]\left[\begin{array}{cc}
I&A^{-1}B\\
0&I
\end{array}\right]\,, \qquad M\equiv F-H-CA^{-1}B.
\eea
Then matrix $\mathfrak{D}$ is given by 
\bea
\mathfrak{D}=\left[\begin{array}{cc}
I&-A^{-1}B\\
0&I
\end{array}\right]\left[\begin{array}{cc}
A^{-1}&0\\
-M^{-1}CA^{-1}&M^{-1}
\end{array}\right],
\eea
in particular,
\bea
\mathfrak{D}_{a\beta}=-[A^{-1}BM^{-1}]_{a\beta}, \quad 
\mathfrak{D}_{\alpha\beta}=[M^{-1}]_{\alpha\beta},\quad 
\mbox{det}\, \mathfrak{D}=[\mbox{det}\, A^{-1}][\mbox{det}\, M^{-1}].
\eea
Recalling that matrices $(A,B,C)$ do not depend on the deformation, we conclude that proving the properties (i)--(iii) amounts to demonstrating than the matrix $M$ does not depend on the coordinates. For example,
equation (\ref{TriangDecomp}) implies that
\bea
\mathfrak{D}(1-P)=S \left[\begin{array}{cc}0&0\\0&M^{-1}\end{array}\right],
\eea
where $S$ does not depend on the deformation and $S_{\alpha\beta}=-\delta_{\alpha\beta}$, so the trivial coordinate dependence of $M$ implies (i) and (ii). 

\bigskip

To summarize, the properties (i)--(iii) would be proven if we demonstrate that $M$ does not depend on coordinates, and this is equivalent to showing that 
\bea
M_0=F-CA^{-1}B
\eea
is a constant matrix. Since the deformation does not enter the last expression, we have arrived at a purely group--theoretic statement, and the rest of this appendix will be dedicated to proving it. 

Let us define $\mathfrak{D}_0$ as the inverse of $(D-\hat{\lambda}^{-1})$ for $H=0$:
\bea\label{DefineD0}
\mathfrak{D}_0=\left[\begin{array}{cc}
D_{ab}-\delta_{ab}&D_{a\beta}\\
D_{\alpha b}&D_{\alpha\beta}
\end{array}\right]^{-1}= \left[\begin{array}{cc}
A&B\\
C&F
\end{array}\right]^{-1}\,.
\eea
Note that $[\mathfrak{D}_0]_{\alpha\beta}=[M_0]_{\alpha\beta}$, and we will show that  these matrix elements do not depend on the coordinates (i.e., on $g$ in (\ref{DabApp})) by demonstrating that they remain constant along {\it any} one--parametric trajectory on a coset. Let us consider such a trajectory:
\bea
g=\exp\left[ixc^\alpha T_\alpha\right]
\eea
Evaluating the derivative of the matrix $D_{AB}$, we find
\bea\label{DerDab}
\frac{d}{dx}D_{AB}=ic^\alpha {f_{B\alpha}}^C D_{AC}
\eea
Introducing a matrix 
\bea
{f_B}^C\equiv c^\alpha {f_{B\alpha}}^C,
\eea
we can solve the differential equation (\ref{DerDab}):
\bea\label{ee1}
D_{AB}(x)={\exp[ix f]_B}^C D_{AC}(0).
\eea
In the canonical gauge (\ref{CanonGauge}) matrix $f$ has only two types of components, ${f_a}^\beta$ and ${f_\alpha}^b$, so we can write\footnote{Due to antisymmetry of the structure constants, matrices $M$ and $N$ are related by $(M\eta)^T=-N\eta$, where $\eta$ is the Killing form. To avoid unnecessary complications, we use canonical generators with $\eta_{AB}=\delta_{AB}$, but obviously the final results (i)--(iii) hold for any normalization, as long as conditions (\ref{CanonGauge}) are satisfied.}
\bea
f=\left[\begin{array}{cc}
0&N^T\\
M^T&0
\end{array}\right]\,,\quad N=-M^T
\eea
and evaluate the exponent
\bea\label{ee2}
\exp[ix f]^T=\left[\begin{array}{cc}
\cos\left[x\sqrt{MN}\right]&ixM\frac{\sin\left[x\sqrt{NM}\right]}{x\sqrt{NM}}\\
ixN\frac{\sin\left[x\sqrt{MN}\right]}{x\sqrt{MN}}&\cos\left[x\sqrt{NM}\right]
\end{array}\right]\,.
\eea
Here we defined two formal functions of matrix variables using series expansions:
\bea\label{MatrSin}
\cos[\sqrt{A}]\equiv\sum_{n=0}^\infty \frac{(-1)^n}{(2n)!}A^n,\qquad
\frac{\sin[\sqrt{A}]}{\sqrt{A}}\equiv\sum_{n=0}^\infty \frac{(-1)^n}{(2n+1)!}A^n\,.
\eea
Matrix $\mathfrak{D}_0$ is determined by substituting (\ref{ee1}) and (\ref{ee2}) into (\ref{DefineD0}).

We begin with analyzing the generic case with $\mbox{det}[MN]\ne 0$. It is natural to identify the starting point $D_{AB}(0)$ of the  trajectory (\ref{ee1}) with the unit element of the group (i.e., with $g=I$ in (\ref{DabApp})), and in our normalization this choice gives\footnote{In general, $D_{AB}$ in the origin is proportional to the Killing form $\eta_{AB}$. To avoid unnecessary complications, we normalized the generators to have $\eta_{AB}=\delta_{AB}$.}
\bea\label{DatZero}
D_{AB}(0)=\delta_{AB}\,.
\eea
Substitution of (\ref{ee1}) and (\ref{ee2}) into (\ref{DefineD0}) with the initial condition (\ref{DatZero}) gives
\bea\label{D0Trig}
\mathfrak{D}_0=\left[\begin{array}{cc}
\cos\left[x\sqrt{MN}\right]-I&ixM\frac{\sin\left[x\sqrt{NM}\right]}{x\sqrt{NM}}\\
ixN\frac{\sin\left[x\sqrt{MN}\right]}{x\sqrt{MN}}&\cos\left[x\sqrt{NM}\right]
\end{array}\right]^{-1}\,.
\eea
Direct calculation shows that, as long as matrices $(MN)$ and $(NM)$ are non--degenerate,  
\bea\label{InvertDnonsing}
\mathfrak{D}_0=\left[\begin{array}{cc}
\cos\left[x\sqrt{MN}\right]&-ixM\frac{\sin\left[x\sqrt{NM}\right]}{x\sqrt{NM}}\\
-ixN\frac{\sin\left[x\sqrt{MN}\right]}{x\sqrt{MN}}&\cos\left[x\sqrt{NM}\right]-I
\end{array}\right]\left[\begin{array}{c}
I-\cos\left[x\sqrt{MN}\right]\qquad   0\quad \\
\qquad 0\qquad  I-\cos\left[x\sqrt{NM}\right]
\end{array}\right]^{-1}\,.
\eea
In particular, it is clear that
\bea\label{D0isOne}
[\mathfrak{D}_0]_{\alpha\beta}=-I
\eea
does not depend on the coordinate $x$. This completes our proof of the statements (i)--(iii) for the trajectories with $\mbox{det}[MN]\ne 0$,  $\mbox{det}[NM]\ne 0$. The rest of this appendix is devoted to the study of degenerate cases. 

First we assume $\mbox{det}[NM]=0$ while still keeping the condition $\mbox{det}[MN]\ne 0$. Then a symmetric matrix $NM$ can be diagonalized by a {\it constant} orthogonal transformation $A$, and after such diagonalization, matrix $M$ can be written in a block form:
\bea
M=\left[\begin{array}{cc}
{\tilde M}&0
\end{array}
\right]A^T\,,\quad \mbox{det}{\tilde M}\ne 0.
\eea
Note that
\bea
N=-A\left[\begin{array}{c}
{\tilde M}^T\\ 0
\end{array}
\right]\,,\quad MN=-{\tilde M}{\tilde M}^T,\quad 
NM=-A\left[\begin{array}{cc}
{\tilde M}^T{\tilde M}&0\\
0&0
\end{array}
\right]A^T\,.
\eea
Substitution into (\ref{D0Trig}) gives
\bea\label{D0TrigA}
\mathfrak{D}_0=
\left[\begin{array}{c|c}
I&0\\
\hline
0&A^T
\end{array}\right]^{-1}
\left[\begin{array}{c|cc}
\cosh\left[x\sqrt{{\tilde M}{\tilde M}^T}\right]-I&ix{\tilde M}
\frac{\sinh\left[x\sqrt{{\tilde M}^T{\tilde M}}\right]}{x\sqrt{{\tilde M}^T{\tilde M}}}&0\\
\hline\\
-ix{\tilde M}^T\frac{\sinh\left[x\sqrt{{\tilde M}{\tilde M}^T}\right]}{x\sqrt{{\tilde M}{\tilde M}^T}}&\cosh\left[x\sqrt{{\tilde M}^T{\tilde M}}\right]&0\\
0&0&I
\end{array}\right]^{-1}
\left[\begin{array}{c|c}
I&0\\
\hline
0&A
\end{array}\right]^{-1}
\,.\nonumber
\eea
Performing the inversion as in (\ref{InvertDnonsing}), we conclude that $(\mathfrak{D}_0)_{\alpha\beta}$ is a constant matrix:
\bea
(\mathfrak{D}_0)_{\alpha\beta}=A\left[\begin{array}{cc}
-I&0\\
0&I
\end{array}\right]A^T\,.
\eea
This completes the proof of the statements (i)--(iii) for all trajectories with $\mbox{det}[MN]\ne 0$.

Finally, we look at the most general case. Diagonalzing symmetric matrices $[MN]$ and $[NM]$ with constant orthogonal rotations $A$ and $B$, we can bring $M$ to a canonical form
\bea
M=B\left[\begin{array}{cc}
{\tilde M}&0\\
0&0
\end{array}
\right]A^T\,,\quad \mbox{det}{\tilde M}\ne 0.
\eea
This gives
\bea
N=-A\left[\begin{array}{cc}
{\tilde M}^T&0\\ 0&0
\end{array}
\right]B^T\,,\ 
MN=-B\left[\begin{array}{cc}
{\tilde M}{\tilde M}^T&0\\
0&0
\end{array}
\right]B^T,\ 
NM=-A\left[\begin{array}{cc}
{\tilde M}^T{\tilde M}&0\\
0&0
\end{array}
\right]A^T\,\nonumber
\eea
and
\bea\label{GeneralGrpElem}
\exp[ix f]^T&=&R\left[\begin{array}{cc|cc}
\cosh\left[x\sqrt{{\tilde M}{\tilde M}^T}\right]&0&ix{\tilde M}
\frac{\sinh\left[x\sqrt{{\tilde M}^T{\tilde M}}\right]}{x\sqrt{{\tilde M}^T{\tilde M}}}&0\\
0&I_{d_1}&0&0\\
\hline
-ix{\tilde M}^T\frac{\sinh\left[x\sqrt{{\tilde M}{\tilde M}^T}\right]}{x\sqrt{{\tilde M}{\tilde M}^T}}&0&\cosh\left[x\sqrt{{\tilde M}^T{\tilde M}}\right]&0\\
0&0&0&I_{d_2}
\end{array}\right]R^{-1},\\
R&=&\left[\begin{array}{c|c}
B&0\\
\hline
0&A\end{array}\right],\quad A^T=A^{-1},\quad B^T=B^{-1}\,\quad \mbox{det}{\tilde M}\ne0.\nonumber
\eea
Substitution of (\ref{GeneralGrpElem}) and (\ref{DatZero}) into (\ref{ee1}) leads to a non--invertible matrix in the right--hand side of (\ref{DefineD0}) unless $d_1=0$. To cure this problem, we observe that under a gauge transformation
\bea
g\rightarrow gh,\quad h\in F,
\eea
matrix (\ref{DabApp}) transforms as
\bea
D_{AB}\rightarrow {{\hat h}_B}~^C D_{AC},
\eea
where ${{\hat h}_B}~^C$ is the image of $h$ in the adjoint representation:
\bea
h T_B h^{-1}\equiv {{\hat h}_B}~^C T_C
\eea
In the basis (\ref{CanonGauge}) matrix ${{\hat h}_B}~^C$ has a block--diagonal form:
\bea
 {{\hat h}_B}~^C=\left[\begin{array}{c|c}
\bullet&0\\
\hline
0&\bullet\end{array}\right]
\eea
To regularize the expression for $\mathfrak{D}_0$ corresponding to (\ref{GeneralGrpElem}), we replace the condition (\ref{DatZero}) by its gauge-transformed version:
\bea
D_{AB}(0)={\hat h}_{BA}.
\eea
Then definition (\ref{DefineD0}) gives
\bea
[\mathfrak{D}_0]^{-1}={\hat h}^TR\left[\begin{array}{cc|cc}
\cosh\left[x\sqrt{{\tilde M}{\tilde M}^T}\right]&0&ix{\tilde M}
\frac{\sinh\left[x\sqrt{{\tilde M}^T{\tilde M}}\right]}{x\sqrt{{\tilde M}^T{\tilde M}}}&0\\
0&I_{d_1}&0&0\\
\hline
-ix{\tilde M}^T\frac{\sinh\left[x\sqrt{{\tilde M}{\tilde M}^T}\right]}{x\sqrt{{\tilde M}{\tilde M}^T}}&0&\cosh\left[x\sqrt{{\tilde M}^T{\tilde M}}\right]&0\\
0&0&0&I_{d_2}
\end{array}\right]R^{-1}-
\left[\begin{array}{c|c}
I&0\\
\hline
0&0\end{array}\right]\nonumber
\eea
Note that the last term in the right--hand side can be written as
\bea
\left[\begin{array}{c|c}
I&0\\
\hline
0&0\end{array}\right]=
{\hat h}^TR\left[\begin{array}{c|c}
{\tilde h}&0\\
\hline
0&0\end{array}\right]R^{-1},
\eea
where ${\tilde h}$ is some matrix. It is convenient to parameterize its components as
\bea
{\tilde h}\equiv \left[\begin{array}{cc}
{\tilde h}_1&{\tilde h}_2\\
{\tilde h}_3&{\tilde h}_4+I_{d_1}\end{array}\right]\,.
\eea
If $d_1$ is even, the we can choose a gauge where ${\tilde h}_2={\tilde h}^T_3=0$, ${\tilde h}_1=I$, and 
\bea
{\tilde h}_4=\exp\left[\begin{array}{cc}0&iq\\ -iq^T&0\end{array}\right]-I_{d_1}
\eea
is a non--degenerate matrix. For odd $d_1$ a similar gauge can be used to reduce the problem to $d_1=1$. Furthermore, by choosing appropriate matrices $A$ and $B$ in (\ref{GeneralGrpElem}), we can make ${\tilde M}$ diagonal, then for $d_1=1$ we can further specify the gauge\footnote{To make the next expression compact, we introduced shortcuts: $\sh=\sinh$, $\ch=\cosh$.}:
\bea
[\mathfrak{D}_0]^{-1}={\hat h}^TR\left[\begin{array}{ccc|ccc}
\ch[x\hat{M}]-I&0&0&i\,\sh[x{\hat M}]&0&0\\
0&\ch [x m]-\ch\, y&i\,\sh\, y&0&i\,\sh[x m]&0\\
0&-i\,\sh\, y&1-\ch\, y&0&0&0\\
\hline
-i\,\sh[x{\hat M}]&0&0&\ch[x{\hat M}]&0&0\\
0&-i\,\sh[x{m}]&0&0&\ch[x m]&0\\
0&0&0&0&0&I_{d_2}
\end{array}\right]R^{-1}\nonumber
\eea
Here ${\hat M}$ is a non--degenerate diagonal matrix, and $m\ne 0$ is a number. 
The inverse of the last matrix is
\bea
\mathfrak{D}_0=R\left[\begin{array}{ccc|ccc}
\frac{\cosh[x\hat{M}]}{\cosh[x\hat{M}]-I}&0&0&i\coth[\frac{x}{2}{\hat M}]&0&0\\
0&\bullet&\bullet&0&\bullet&0\\
0&\bullet&\bullet&0&\bullet&0\\
\hline
-i\coth[\frac{x}{2}{\hat M}]&0&0&-I&0&0\\
0&\bullet&\bullet&0&1&0\\
0&0&0&0&0&I_{d_2}
\end{array}\right][{\hat h}^TR]^{-1}\nonumber
\eea
Bullets denote some complicated expressions which are irrelevant for our analysis. 

To summarize, we have demonstrated that even in the degenerate case when $\mbox{det}[MN]= 0$, there exists a gauge where $[\mathfrak{D}_0]_{\alpha\beta}$ remains constant along any one--parametric trajectory. This completes the proof of the statements (i)--(iii).

\end{document}